\documentclass[11pt]{article}
\usepackage{subfigure,graphicx}

\newcommand{\BABARPubYear}    {06}

\newcommand{\BABARConfNumber} {038}
\newcommand{\SLACPubNumber} {12028}

\newcommand{\reslinepolvalppp}[4]{\left(#1\pm#2\pm#3\pm#4\right)^\circ}

\input pubboard/babarsym
\def\Dztilde   {\ensuremath {\tilde{D}^0}\xspace}
\def\bea{\begin{eqnarray}}
\def\eea{\end{eqnarray}}

\def\dodstartilde {\ensuremath {\tilde{D}^{(\ast)0}}\xspace}
\def\K  {\ensuremath{K}\xspace}

\newcommand{\npa}       [1]  {\npBase\ A~{\bf #1}}

\def\bea{\begin{eqnarray}}
\def\eea{\end{eqnarray}}

\newcommand{\re}{\ensuremath{\mathop{\rm Re}}}
\newcommand{\im}{\ensuremath{\mathop{\rm Im}}}

\newcommand{\fis}{\ensuremath{\mbox{$\mathcal{F}$}}\xspace}

\long\def\inst#1{\par\nobreak\kern 4pt\nobreak
    {\it #1}\par\vskip 10pt plus 3pt minus 3pt}

\begin{document}
{\pagestyle{empty}


\begin{flushright}
\babar-CONF-\BABARPubYear/\BABARConfNumber \\
SLAC-PUB-\SLACPubNumber \\
\end{flushright}

\par\vskip 5cm

\begin{center}
\Large \bf Measurement of the CKM angle $\gamma$ in $B^\mp\rightarrow D^{(*)}K^\mp$ decays with a Dalitz analysis of $D^0\rightarrow K^0_s\pi^-\pi^+$
\end{center}
\bigskip

\begin{center}
\large The \babar\ Collaboration\\
\mbox{ }\\
\today
\end{center}
\bigskip \bigskip

\begin{center}
\large \bf Abstract
\end{center}
We present a measurement of the Cabibbo-Kobayashi-Maskawa $CP$-violating phase $\gamma$ with a Dalitz analysis of neutral $D$-meson decays to the $K^0_s\pi^-\pi^+$ final state from $B^\mp\rightarrow D^{(*)}K^\mp$ decays, using a sample of $347$ million $B\bar{B}$ events collected by the \babar\ detector. 
We measure $\gamma = \reslinepolvalppp{92}{41}{11}{12}$, where the first error is statistical, 
the second is the experimental systematic uncertainty and the third reflects the Dalitz model uncertainty. 
For the ratios $r_B^{(*)}$ between the magnitudes of amplitudes ${\cal A} (\Bm\to D^{(*)0} \Km)$ and ${\cal A} (\Bm\to \bar{D}^{(*)0} \Km)$
we obtain the one-standard deviation intervals $[0,0.14]$ and $[0.02,0.20]$, respectively. All results presented here are preliminary.
\vfill
\begin{center}

Submitted to the 33$^{\rm rd}$ International Conference on High-Energy Physics, ICHEP 06,\\
26 July---2 August 2006, Moscow, Russia.

\end{center}

\vspace{1.0cm}
\begin{center}
{\em Stanford Linear Accelerator Center, Stanford University, 
Stanford, CA 94309} \\ \vspace{0.1cm}\hrule\vspace{0.1cm}
Work supported in part by Department of Energy contract DE-AC03-76SF00515.
\end{center}

\newpage
} 

\begin{center}
\small

The \babar\ Collaboration,
\bigskip

%
{B.~Aubert,}
{R.~Barate,}
{M.~Bona,}
{D.~Boutigny,}
{F.~Couderc,}
{Y.~Karyotakis,}
{J.~P.~Lees,}
{V.~Poireau,}
{V.~Tisserand,}
{A.~Zghiche}
\inst{Laboratoire de Physique des Particules, IN2P3/CNRS et Universit\'e de Savoie,
 F-74941 Annecy-Le-Vieux, France }
% ...

\end{center}\newpage


\section{INTRODUCTION}\label{sec:intro}

\begin{figure}[!h]
  \begin{center}
\includegraphics[height=5cm]{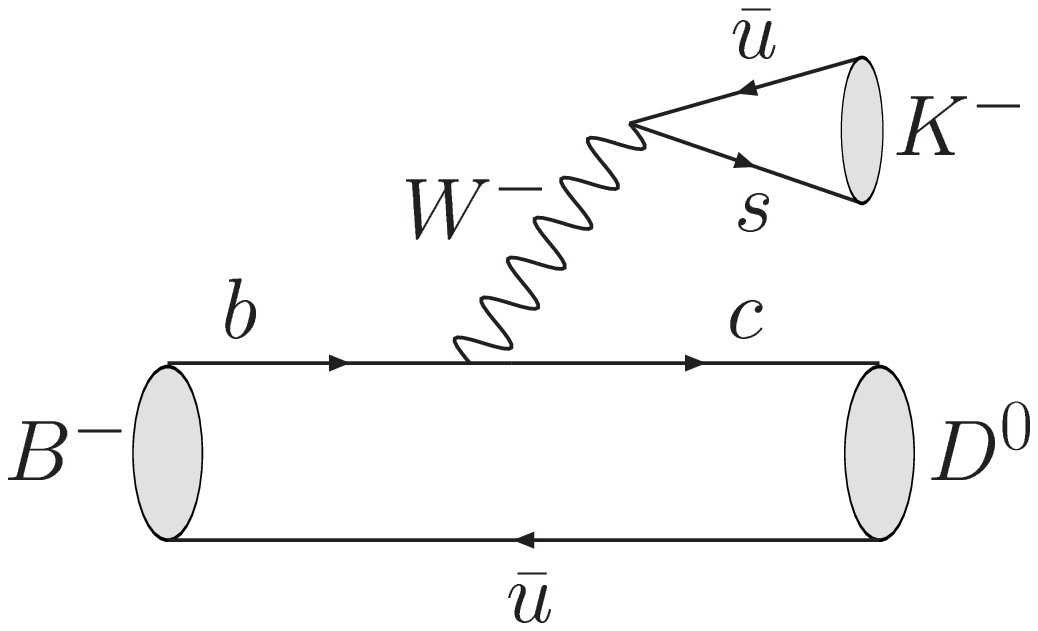}
    \hspace{1cm}
    \includegraphics[height=5cm]{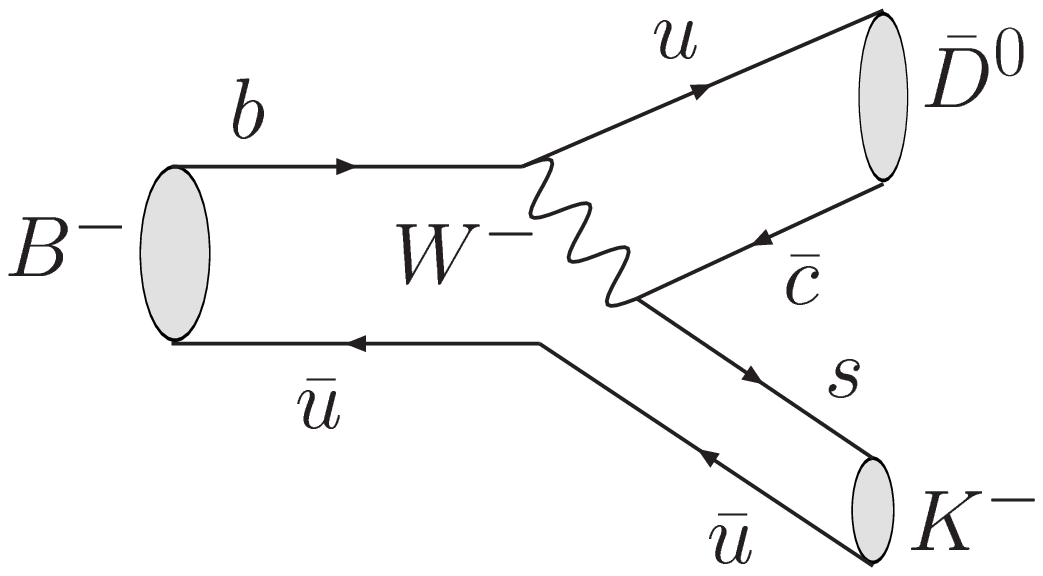}
    \caption{ Diagrams contributing to $B^- \to \Dztilde K^-$ decay. The 
      left diagram proceeds via $\b \to \c\ubar\s$ transition, while the right diagram proceeds via 
       $\b \to \u \cbar \s$ transition and is color suppressed.}
    \label{fig:DK_diag}
    \end{center}
  \end{figure}

The angle $\gamma$ of the unitarity triangle is the phase of the Cabibbo-Kobayashi-Maskawa (CKM)
matrix~\cite{ref:cabibbo} defined as $\gamma\equiv\arg{\left[-V_{ud}^{}V_{ub}^{*}/V_{cd}^{}V_{cb}^{*}\,\right]}$,
which corresponds to the phase of the element $V^*_{ub}$, i.e. $V_{ub} = |V_{ub}|e^{-i\gamma}$,
in the Wolfenstein parameterization~\cite{ref:wolfenstein}.
Various methods have been proposed to 
extract $\gamma$ using $B^\mp\to \Dztilde K^\mp$ decays, all exploiting the
interference between the color allowed $B^-\to D^0K^-$ ($\b \to \c\ubar\s \propto
V_{cb}$) and the color suppressed $B^-\to \bar{D^0}K^-$ ($\b \to \u \cbar \s \propto
V_{ub}$) transitions~\cite{ref:chargeconj}, when the $D^0$ and $\bar{D^0}$ are reconstructed
in a common final state~\cite{ref:gronau,ref:soni,ref:ggsz_ads,ref:belle_dal04}. The symbol \Dztilde indicates either a $D^0$ or
a $\bar{D}^0$ meson. The extraction of $\gamma$ with these decays
is theoretically clean because the main contributions to the
amplitudes come from tree-level diagrams (see Fig.~\ref{fig:DK_diag}).

Both \babar~\cite{ref:babar_dalitzpub,ref:babar_dalitzeps05} and 
Belle~\cite{ref:belle_dalitz} have reported on a measurement of $\gamma$ based
on $\Bm \to \dodstartilde K^-$ and $\Bm\to \Dztilde K^{*-}$ decays with a Dalitz analysis of
$\Dztilde \to \KS \pim \pip$, with $D^{*0}\rightarrow D^0\pi^0$ and
$D^{*0}\rightarrow D^0\gamma$ (\babar\ only), and $K^{*-}\rightarrow
\KS \pi^-$. 
In this paper we report on an update 
with $\Bm \to \tilde{D}^{(*)0}K^-$ decays.

Assuming no \CP asymmetry in $D^0$ decays, the $\Bmp \to \tilde{D}^{(*)0} \Kmp$, $\tilde{D}^{*0} \rightarrow \Dztilde\pi^0,\Dztilde\gamma$,
$\Dztilde \to \KS \pim \pip$ decay chain rate
$\Gamma^{(*)}_\mp(m^2_-,m^2_+)$ can be \mbox{written
  as}~\cite{ref:ggsz_ads}
\bea
\Gamma^{(*)}_\mp(m^2_-,m^2_+) & \propto & |{\cal A}_{D\mp}|^2 + {r^{(*)}_B}^2 |{\cal A}_{D\pm}|^2 + \nonumber \\ 
& & 2 k r^{(*)}_B\,\left\{ \cos(\delta^{(*)}_B \mp\gamma) \re[{\cal A}_{D\mp} {\cal A}_{D\pm}^*]+ \sin(\delta^{(*)}_B \mp\gamma) 
       \im[ {\cal A}_{D\mp} {\cal A}^*_{D\pm}] \right\} ~,
\label{eq:ampgen1}
\eea
where $m^2_-$ and $m^2_+$ are the squared invariant masses of the
$\KS\pim$ and $\KS\pip$ combinations, respectively,
and ${\cal A}_{D\mp} \equiv {\cal A}_{D}(m^2_\mp,m^2_\pm)$,
with ${\cal A}_{D-}$ (${\cal A}_{D+}$) the amplitude of the $\Dz \to
\KS\pim\pip$ ($\Dzb \to \KS\pip\pim$) decay.
The value of the \CP-odd phase $\gamma$ changes sign for \Bp\ and \Bm in Eq.~(\ref{eq:ampgen1}),  
leading to different rates in corresponding regions of the \Dz Dalitz plane, for \Bp\ and \Bm\ decays.
We introduce here the {\it \CP (cartesian) parameters} $x^{(*)}_\mp$ and $y^{(*)}_\mp$~\cite{ref:babar_dalitzpub}, 
defined respectively as the real and imaginary part of $r^{(*)}_B e^{i(\delta^{(*)}_B \mp\gamma)}$, for which the
constraint ${r^{(*)}_B}^2={x^{(*)}_\mp}^2 +{y^{(*)}_\mp}^2$ holds. 
Here, $r^{(*)}_B$ is the magnitude
of the ratio of the amplitudes ${\cal A}(\Bm \to \bar{D}^{(*)0} \Km)$ and
${\cal A}(\Bm \to D^{(*)0} \Km)$ and $\delta^{(*)}_B$ is their relative strong
phase. As a consequence of parity and angular momentum conservation in
the \dodstartilde decay, the factor $k$ in Eq.~(\ref{eq:ampgen1}) takes the value $+1$ for $\Bmp
\to \Dztilde \Kmp$ and $\Bmp \to \tilde{D}^{*0}(\Dztilde\pi^0) \Kmp$, and
$-1$ for $\Bmp \to \tilde{D}^{*0}(\Dztilde\gamma)
\Kmp$~\cite{ref:bondar_gershon}.

Once the decay amplitude ${\cal A}_{D}$ is known, the Dalitz plot distributions 
for \Dztilde from $\B^- \to \tilde{D}^{(*)0} \K^-$ and  $\B^+ \to \tilde{D}^{(*)0} \K^+$ decays can be simultaneously 
fitted to $\Gamma^{(*)}_-(m^2_-,m^2_+)$ and $\Gamma^{(*)}_+(m^2_-,m^2_+)$ as given by Eq.~(\ref{eq:ampgen1}), 
respectively. 
A maximum likelihood technique is used to measure
the \CP-violating parameters $x_\mp^{(*)}$, $y_\mp^{(*)}$.
From them, confidence regions for $\gamma$, $r^{(*)}_B$ and $\delta^{(*)}_B$ are obtained with a frequentist method.
We extract $x^{(*)}_\mp$, $y^{(*)}_\mp$ instead of
$\gamma,\delta_B^{(*)},r_B^{(*)}$ because the distributions of the
cartesian parameters are unbiased and 
Gaussian, while the
distributions of $\gamma,\delta_B^{(*)},r_B^{(*)}$ don't have these properties for small values of $r_B^{(*)}$ and low-statistics samples.

\section{THE \babar\ DETECTOR AND DATASET}\label{sec:dataset}
The analysis is based on a sample of approximately $347$ million \BB pairs collected by the \babar\ detector at the SLAC PEP-II $e^+e^-$ asymmetric-energy storage ring. The \babar\ detector is optimized for the asymmetric-energy beams at PEP-II and is described in~\cite{ref:babar}. We summarize briefly the components that are crucial to this analysis. Charged-particle tracking is provided by a five-layer silicon vertex tracker (SVT) and a 40-layer drift chamber (DCH). In addition to providing precise space coordinates for tracking, the SVT and DCH also measure the specific ionization ($dE/dx$), which is used for particle identification of low-momentum charged particles. At higher momenta ($p>0.7$~\gevc)
pions and kaons are identified by Cherenkov radiation detected in a ring-imaging
device (DIRC). The typical separation between pions and kaons varies from 8$\sigma$
at 2~\gevc to 2.5$\sigma$ at 4~\gevc. 
The position and energy of photons are
measured with an electromagnetic calorimeter (EMC) consisting of 6580 thallium-doped CsI crystals.
These systems are mounted inside a 1.5 T solenoidal super-conducting magnet. 

\section{EVENT SELECTION}\label{sec:selection}
We reconstruct the $\Bm \to \dodstartilde K^-$ decays
with $\dodstartilde \rightarrow \Dztilde\pi^0,\Dztilde\gamma$ and 
$\Dztilde\to \KS \pi^-\pi^+$~\cite{ref:chargeconj}. 
The \KS candidates are formed from oppositely charged pions
with a reconstructed invariant mass within 9~\mevcc of the nominal \KS
mass~\cite{ref:pdg2004}. The two pions are constrained to originate
from the same point. The $\Dztilde\to\KS \pi^-\pi^+$ candidates are selected by combining mass constrained \KS
candidates with two oppositely charged pions having an
invariant mass within 12~\mevcc of the nominal $D^0$
mass~\cite{ref:pdg2004}. 
The $\pi^0$ candidates from $D^{*0}\to D^0\pi^0$ are formed from pairs
of photons with invariant mass in the range $[115,150]$~\mevcc, and
with photon energy greater than 30~\mev. Photon candidates from
$D^{*0}\to D^0\gamma$ are selected if their energy is greater than
100~\mev. $D^{*0}\to D^0\pi^0 (D^0\gamma)$ candidates are required to
have a $D^{*0}$-$D^0$ mass difference within 2.5~(10)~\mevcc of its
nominal value~\cite{ref:pdg2004}, corresponding to about two standard deviations.
$B^-\to\dodstartilde K^-$ candidates are formed by combining a
\dodstartilde candidate with a track identified as a kaon. 

We select $B$ mesons by using the energy
difference $\Delta E=E^*_B-E^*_i/2$, and the beam-energy substituted mass, $\mes=
\sqrt{(E^{*2}_i/2+{\bf p_i\cdot p_B})^2/E^2_i-p^2_B}$, 
where the subscripts $i$ and $B$
refer to the initial $e^+e^-$ system and the $B$ candidate,
respectively, and the asterisk denotes the center-of-mass (CM)
frame. 
The resolution of \DeltaE
ranges between $15\mev$ and $18\mev$ depending on the decay mode.
The resolution of \mes 
is about 2.6~\mevcc for all the $B$ decay modes.
We define a selection region through the requirement $-80<\DeltaE<120$~\mev and
$\mes>5.2$~\gevcc. 
To suppress $e^+e^-\to q\bar{q}$,  $q=u,d,s,c$ (continuum) events, we
require $|\cos\theta_T|< 0.8$ where $\theta_T$ is defined as the angle
between the thrust axis of the \B candidate and that of the rest of
the event. Furthermore we define a Fisher discriminant \fis\ that we use in a likelihood fit 
to separate continuum and \BB events. It is defined as a linear combination 
of four topological variables: $L_0=\sum_{i}
p_i^*$, $L_2 =\sum_{i} p_i^*  |\cos \theta^*_i|^2$, the absolute 
value of the cosine of the CM polar angle of the \B candidate momentum, 
and $|\cos\theta_T|$. Here, $p_i^*$ and $\theta_i^*$ are the
CM momentum and the angle of the remaining tracks and clusters in the event, with respect to the \B candidate thrust axis.
If both $B^-\to \tilde{D}^{*0}(\Dztilde\pi^0)K^-$ and $B^-\to \tilde{D}^{*0}(\Dztilde\gamma)K^-$ candidates are selected in the same event, only the $B^-\to \tilde{D}^{*0}(\Dztilde\pi^0)K^-$ is kept. The cross-feed among the different samples is negligible except for $B^-\to \tilde{D}^{*0}(\Dztilde\gamma)K^-$, where the background from $B^-\to \tilde{D}^{*0}(\Dztilde\pi^0)K^-$ is about 5\% of the signal yield. This contamination has a negligible effect on the measurement of the \CP parameters.

\begin{figure}[htb!]
\begin{center}
\begin{tabular}{cc}
\includegraphics[height=7cm]{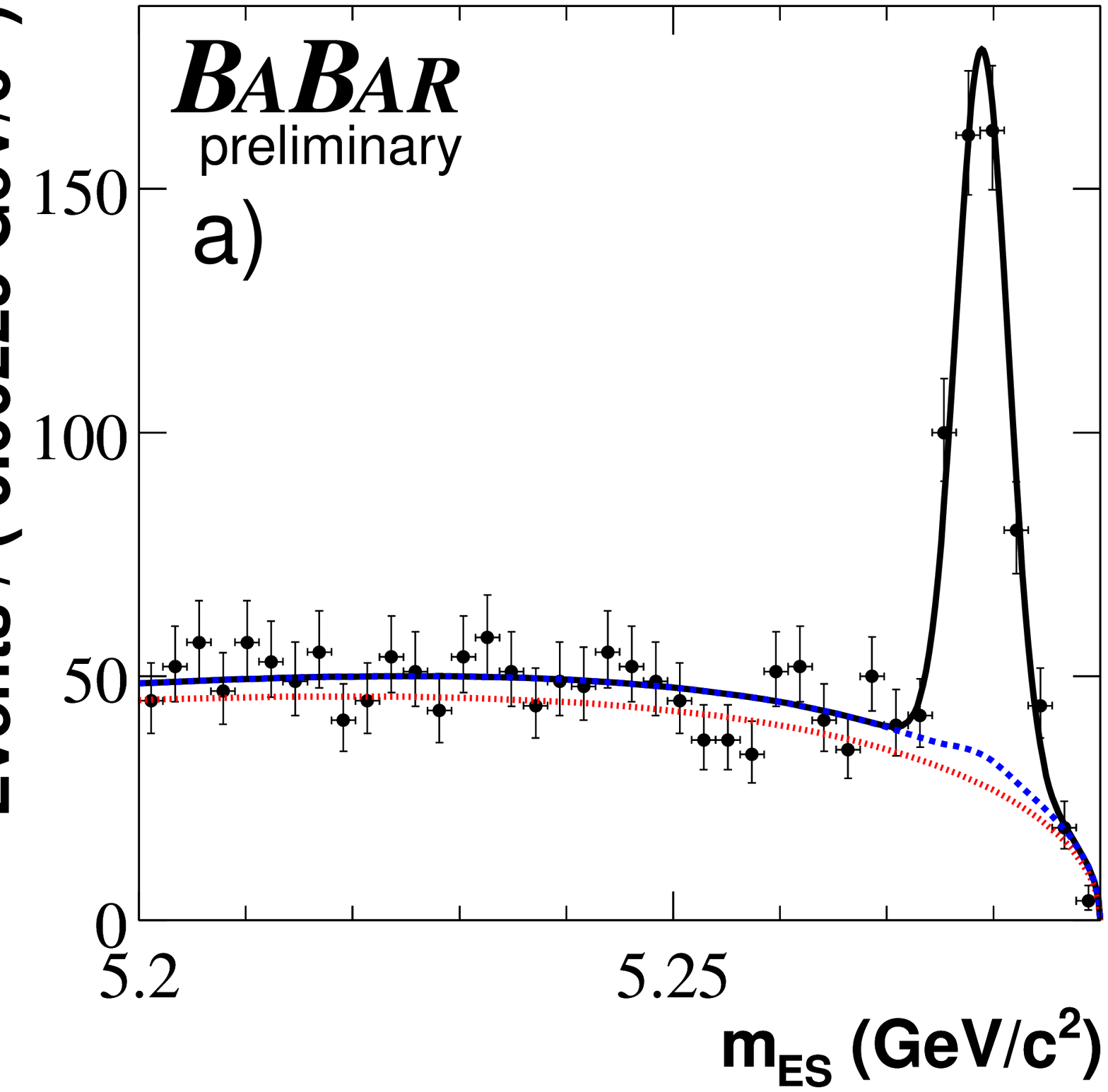} \\
\includegraphics[height=7cm]{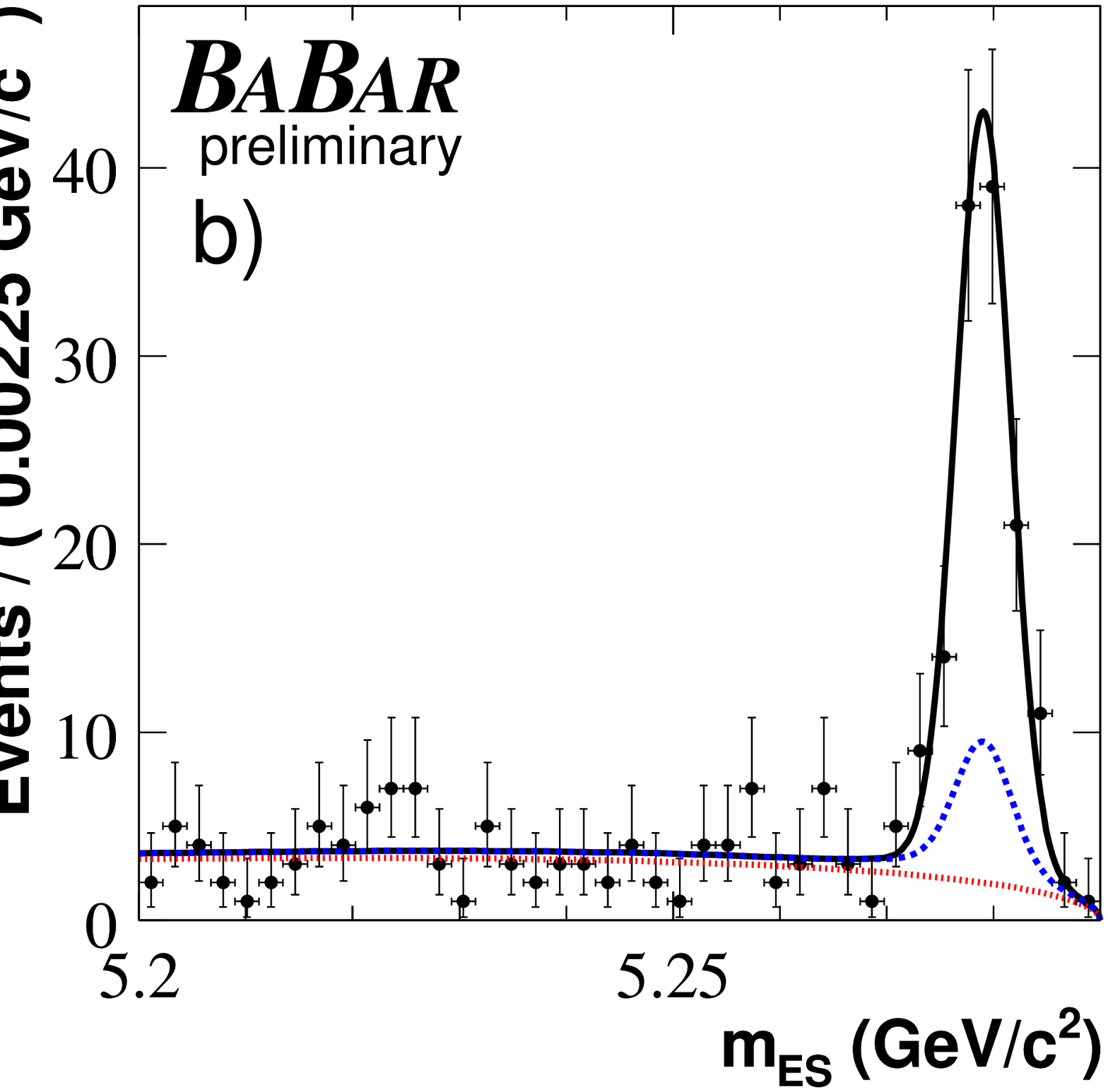}
\includegraphics[height=7cm]{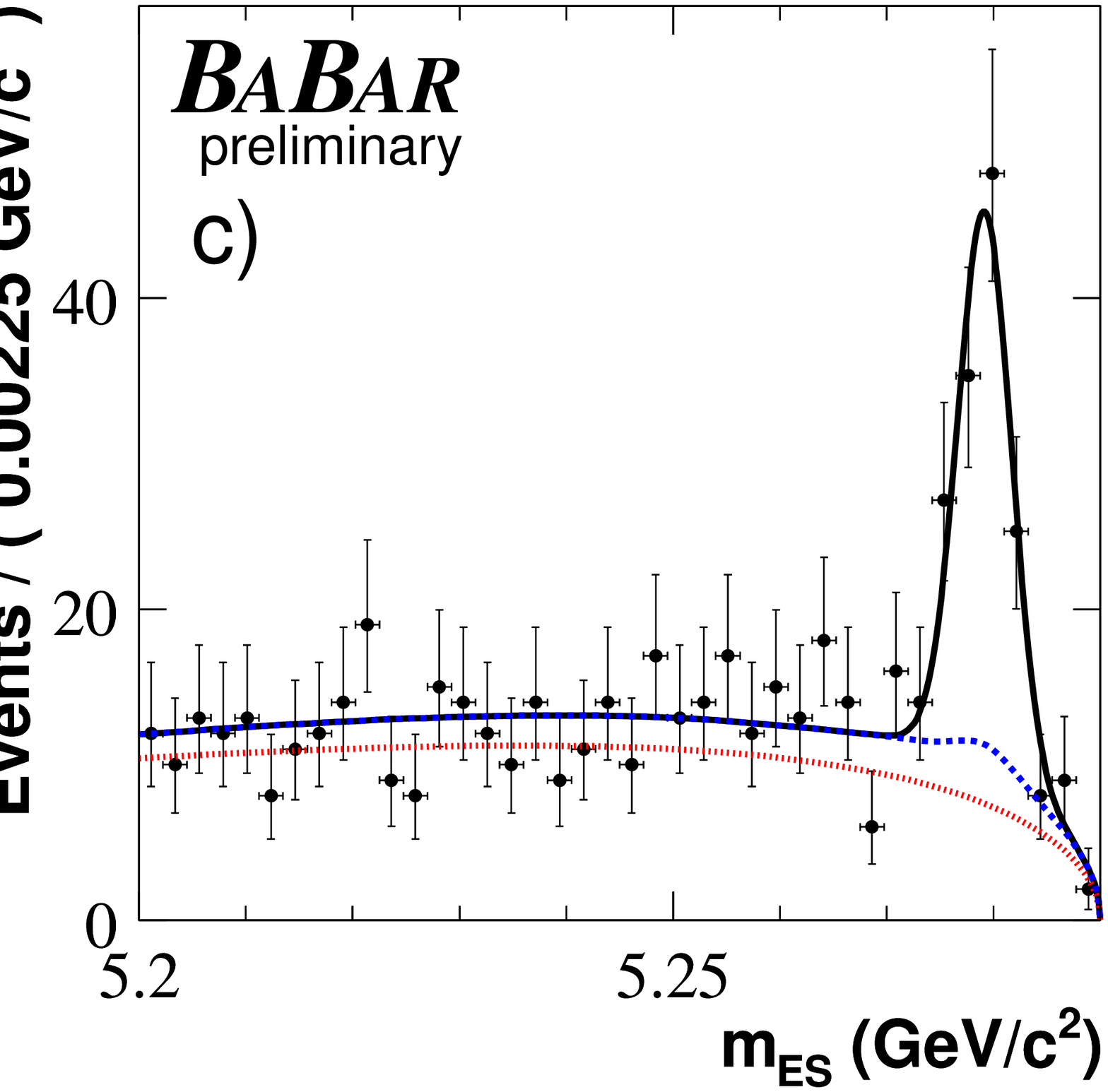}
\end{tabular}
\caption{Distributions of \mes\ for (a) $B^-\to \Dztilde K^-$, 
(b) $B^-\to \tilde{D}^{*0}(\Dztilde\pi^0)K^-$, and (c) $B^-\to \tilde{D}^{*0}(\Dztilde\gamma)K^-$.
The curves superimposed represent the overall fit projections (solid black lines), the continuum contribution (dotted red lines), 
and the sum of all background components (dashed blue lines).
}
\label{fig:btodk_mes}
\end{center}
\end{figure}

The reconstruction efficiencies are $15\%$, $7\%$, $9\%$, 
for the $B^-\to \Dztilde K^-$, $B^-\to
\tilde{D}^{*0}(\Dztilde\pi^0)K^-$ and $B^-\to
\tilde{D}^{*0}(\Dztilde\gamma)K^-$ decay modes, respectively. 
Fig.~\ref{fig:btodk_mes} shows the \mes\ distributions after all selection criteria plus 
a tighter requirement on \DeltaE, $|\DeltaE|<30$~\mev, are applied.
The largest background contribution is from continuum events or \BB decays where a fake or true $D^0$ is combined with a random track. Another source of background is given by those $B^-\to D^{(*)0}\pi^-$ decays where the prompt pion is misidentified as kaon. These decays are separated from the signal 
using their different \DeltaE distribution.

\section{The $D^0\rightarrow \KS \pi^-\pi^+$  DECAY MODEL}\label{sec:dalitzmodel}

\begin{figure}[!ht]
\begin{center}
\begin{tabular} {cc}  
{\includegraphics[height=6cm]{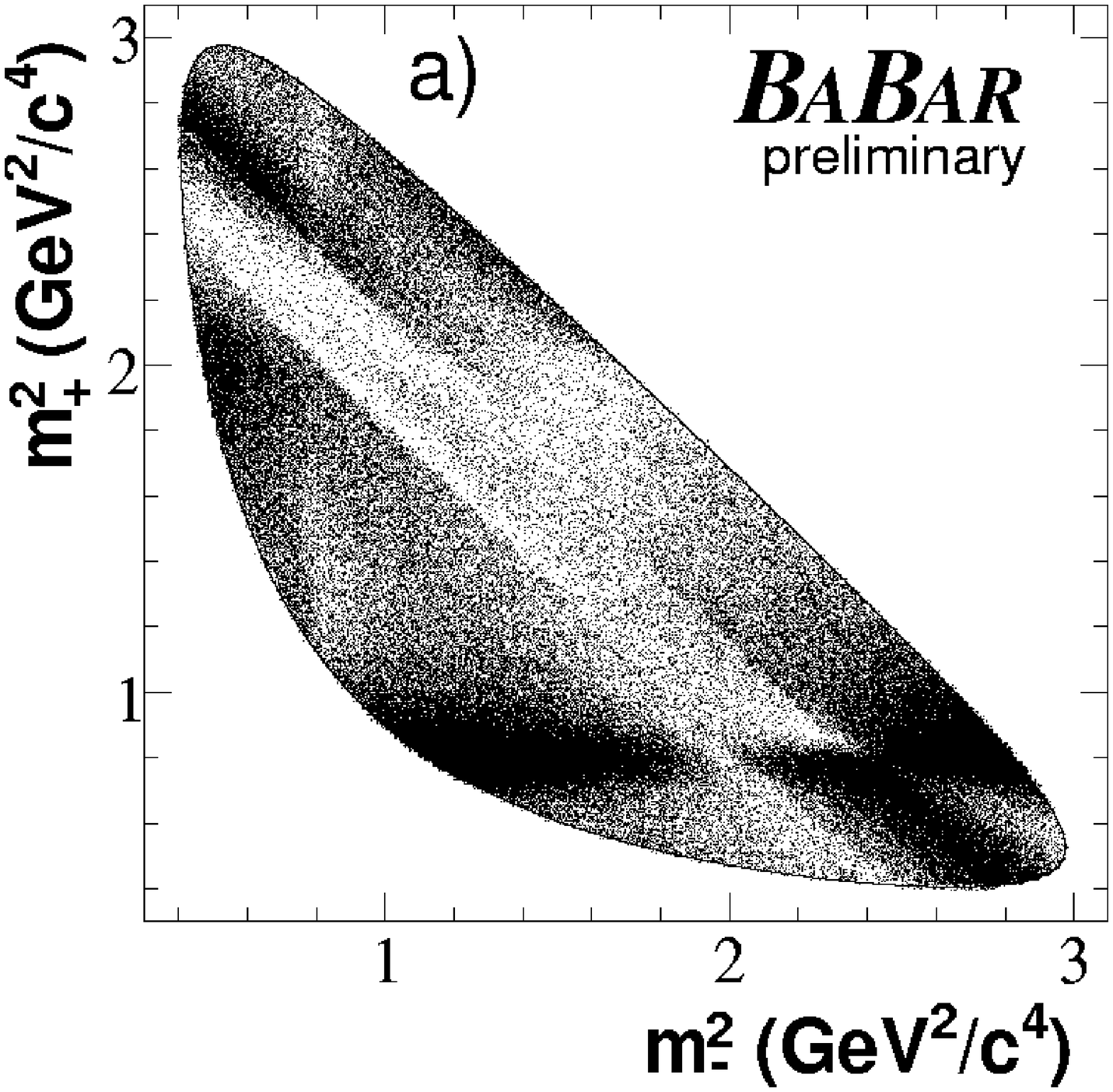}} &
{\includegraphics[height=6cm]{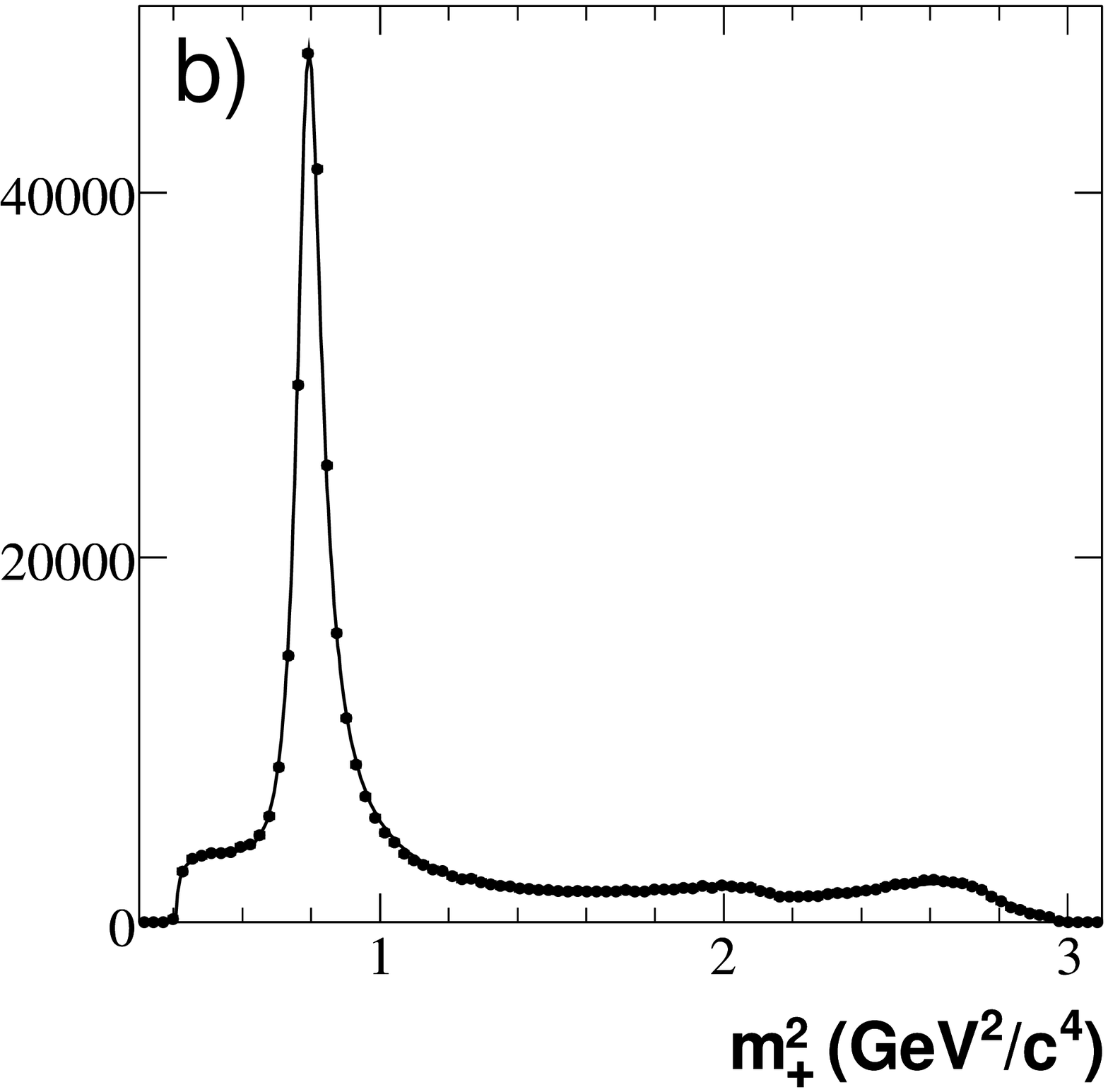}} \\
{\includegraphics[height=6cm]{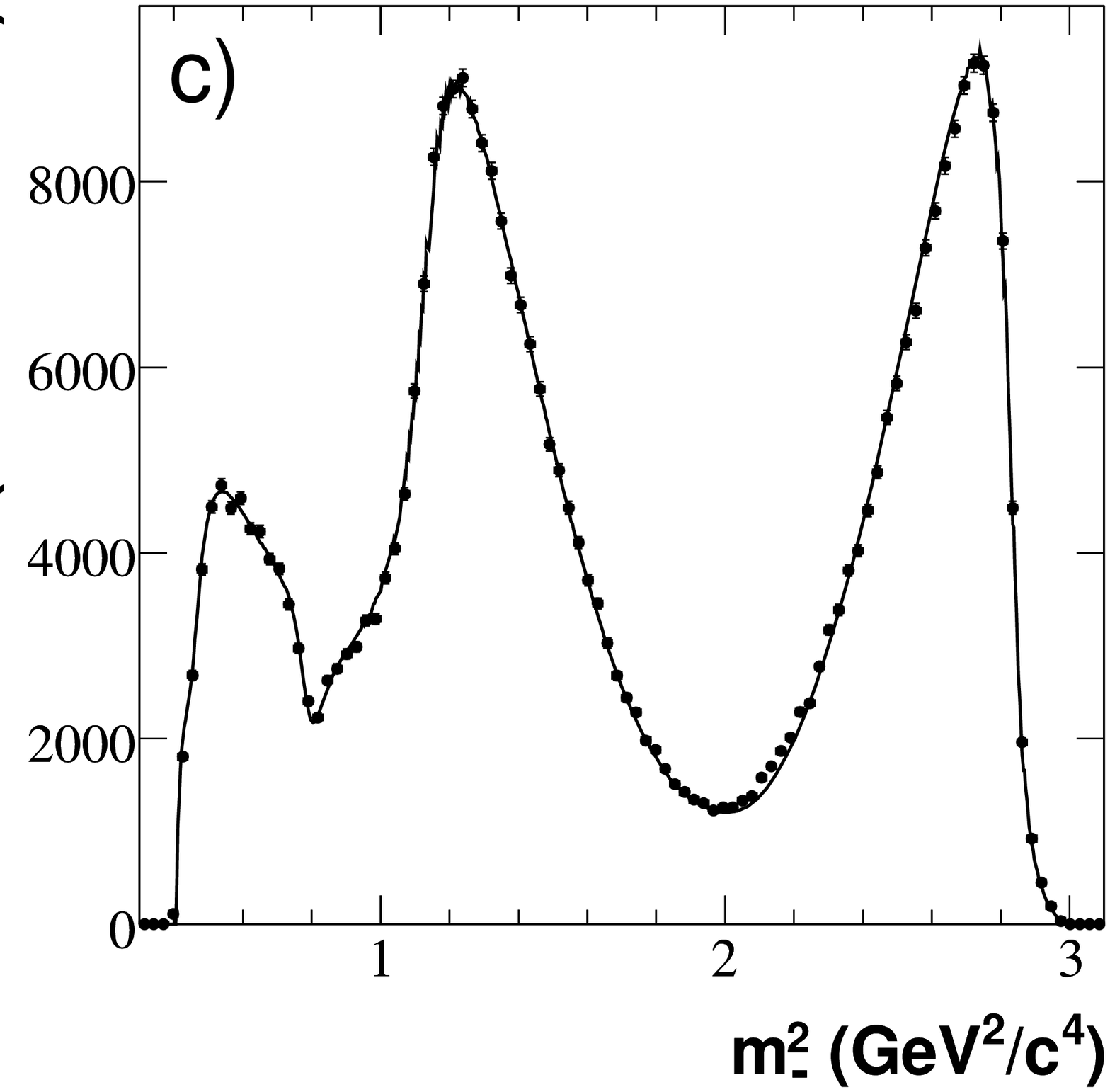}} &
{\includegraphics[height=6cm]{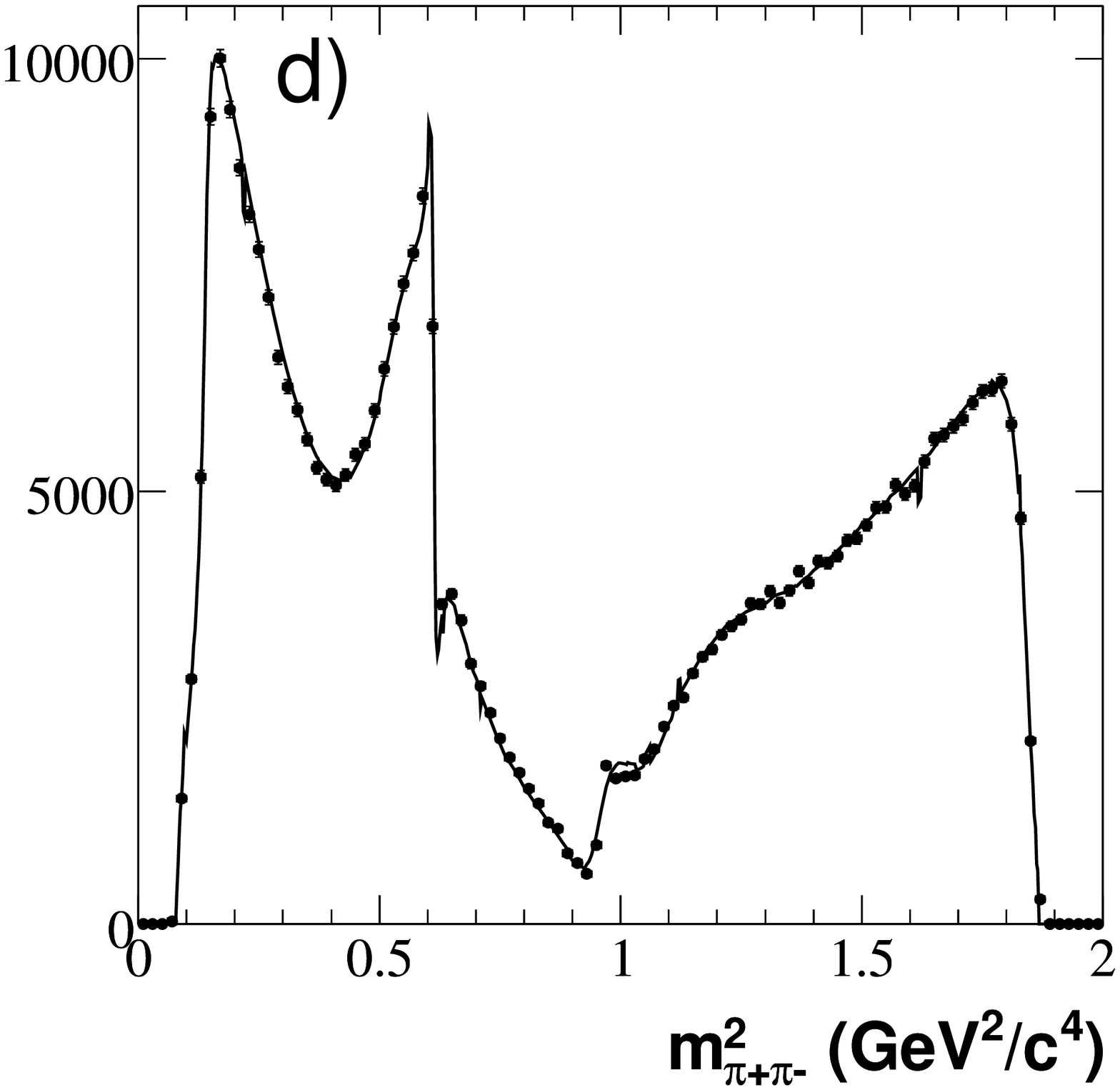}} \\
\end{tabular}   
\caption{(a) The $\bar{D}^0 \to \KS \pim \pip$ Dalitz distribution from $D^{*-} \to \bar{D}^0 \pi^-$ events, and projections
on (b) $m^2_+=m^2_{\KS\pi^+}$, (c) $m^2_-=m^2_{\KS\pi^-}$, and (d) $m^2_{\pip\pim}$. $\Dz \to \KS \pi^+ \pi^-$ from $\Dstarp \to \Dz \pip$ events 
are also included. The curves are the reference model fit projections. 
}
\label{fig:BWfit-res}
\end{center}
\end{figure}

The $\Dz \to \KS \pim \pip$ decay amplitude ${\cal A}_D(m^2_-,m^2_+)$ is determined from an unbinned maximum-likelihood fit to
the Dalitz plot distribution of a high-purity (97.7\%) $D^0$ sample from 390328 $\Dstarp\to\Dz\pip$ decays reconstructed in 270 \invfb of data, 
shown in Fig.~\ref{fig:BWfit-res}. 
Our reference model to describe ${\cal A}_D(m^2_-,m^2_+)$ is based on Breit-Wigner (BW) parameterizations of a set of resonances, and 
is the same as used for our previously reported measurement 
of $\gamma$ on $\Bm \to \dodstartilde \Km$, $\Bm \to \Dztilde \Kstarm$, $\Dztilde \to \KS \pim \pip$ decays~\cite{ref:babar_dalitzpub,ref:babar_dalitzeps05}.

The decay amplitude in the reference model is expressed as a sum of 
two-body decay-matrix elements (subscript $r$) and a non-resonant (subscript NR) contribution,
\bea
{\cal A}_D(m^2_-,m^2_+) = \Sigma_r a_r e^{i \phi_r} {\cal A}_r(m^2_-,m^2_+) + a_{\rm NR} e^{i \phi_{\rm NR}}~,
\eea
where each term is parameterized with an amplitude $a_r$ ($a_{\rm NR}$)
and a phase $\phi_r$ ($\phi_{\rm NR}$).
The function ${\cal A}_r(m^2_-,m^2_+)$ is the Lorentz-invariant expression for the matrix element of a \Dz meson 
decaying into $\KS \pim \pip$ through an intermediate resonance $r$, parameterized as a
function of position in the Dalitz plane. 
For $r=\rho(770)$ and $\rho(1450)$ we use the functional form 
suggested in Ref.~\cite{ref:gounarissakurai}, while the remaining resonances 
are parameterized by a spin-dependent relativistic BW distribution. The angular dependence of the BW terms is described with 
the helicity formalism as shown in \cite{ref:cleo}\footnote{The label A and B should be swapped in Eq. (6) of \cite{ref:cleo}.}. 
Mass and width values  are taken from~\cite{ref:pdg2004}, with the exception of $K^{*}_0(1430)^+$ taken from \cite{ref:e791K*}.  
The model consists of 13 resonances leading to 16 two-body decay amplitudes and phases 
(see Table~\ref{tab:BWfit-res}), plus the non-resonant contribution, and accounts
for efficiency variations across the Dalitz plane and the small background contribution.
All the  resonances considered in this model are well established except for the two 
scalar $\pi\pi$  resonances, $\sigma$ and $\sigma'$, whose
masses and widths are obtained from our sample~\cite{ref:comment_sigma}. 
Their addition to the model is motivated by an improvement in the description of the data. 

The possible absence of the
$\sigma$ and $\sigma'$ resonances is considered in the evaluation of
the systematic errors.
In this respect, the K-matrix formalism~\cite{ref:Kmatrix} provides a direct way of imposing
the unitarity constraint that is not guaranteed in the case of the BW model and is 
suited to the study of broad and overlapping resonances in multi-channel decays.
We use the K-matrix method to parameterize
the $\pi\pi$ S-wave states, avoiding the need to introduce the two
$\sigma$ scalars. A description of this alternative model can be found in \cite{ref:babar_dalitzeps05}.

\begin{table}[h]
\begin{center}
\begin{tabular}{l|c|c|c}
\hline
\\[-0.15in]
    Component  &  $Re\{a_r e^{i\phi_r}\}$ &  $Im\{a_r e^{i\phi_r}\}$ & Fit fraction (\%) \\ [0.01in]
\hline \hline
$K^{*}(892)^-$        &  $-1.223\pm0.011$  &  $1.3461\pm0.0096$  &  58.1 \\  
$K^{*}_0(1430)^-$     &    $-1.698\pm0.022$   &   $-0.576\pm0.024$  & 6.7 \\  
$K^{*}_2(1430)^-$     &    $-0.834\pm0.021$    &  $0.931\pm0.022$  &  3.6 \\  
$K^{*}(1410)^-$       &   $-0.248\pm0.038$  &   $-0.108\pm0.031$    & 0.1 \\ 
$K^{*}(1680)^-$       &    $-1.285\pm0.014$     &   $0.205\pm0.013$    &  0.6 \\ 
\hline
$K^{*}(892)^+$        &   $0.0997\pm0.0036$   & $-0.1271\pm0.0034$   & 0.5 \\ 
$K^{*}_0(1430)^+$     &    $-0.027\pm0.016$    &    $-0.076\pm0.017$   & 0.0 \\ 
$K^{*}_2(1430)^+$     &  $0.019\pm0.017$   &  $0.177\pm0.018$   &  0.1 \\ 
\hline
$\rho(770)$           &     $1$     &      $0$                    &   21.6 \\ 
$\omega(782)$         &  $-0.02194\pm0.00099$  & $0.03942\pm0.00066$   & 0.7 \\ 
$f_2(1270) $          &    $-0.699\pm0.018$   &   $0.387\pm0.018$   & 2.1 \\ 
$\rho(1450)$          &    $0.253\pm0.038$    &   $0.036\pm0.055$    &   0.1 \\ 
\hline 
Non-resonant          &    $-0.99\pm0.19$   &     $3.82\pm0.13$    &    8.5 \\ 
$f_0(980) $           &   $0.4465\pm0.0057$   &   $0.2572\pm0.0081$   &  6.4 \\ 
$f_0(1370) $          &   $0.95\pm0.11$   &    $-1.619\pm0.011$      &  2.0 \\ 
$\sigma$              &    $1.28\pm0.02$   &  $0.273\pm0.024$    &   7.6 \\ 
$\sigma '$            &  $0.290\pm0.010$   &   $-0.0655\pm0.0098$    &   0.9 \\ 
\hline
\end{tabular}
\end{center}
\caption{Complex amplitudes $a_r e^{i\phi_r}$ and fit fractions of the different components ($K_S\pi^-$, 
$K_S\pi^+$, and $\pi^+\pi^-$ resonances) obtained from the fit of the $D^0 \to K_S\pi^-\pi^+$ Dalitz 
distribution from $D^{*+} \to D^0 \pi^+$ events. Errors are statistical only. Masses and widths of all 
resonances are taken from~\cite{ref:pdg2004} with the exception of $K^{*}_0(1430)^+$ taken from~\cite{ref:e791K*}. 
The fit fraction is defined for the resonance terms as 
the integral of $a_r^2 |{\cal A}_r(m^2_-,m^2_+)|^2$ over the Dalitz plane divided by the integral
of $|{\cal A}_D(m^2_-,m^2_+)|^2$. The sum of fit fractions is
$119.5\%$. A value different from 100\% is a consequence of the interference among the amplitudes.
}
\label{tab:BWfit-res}
\end{table}

\section{$CP$ ANALYSIS}\label{sec:cpfit}
We simultaneously fit the $\B^\mp \to \tilde{D}^{(*)0} \K^\mp$ 
samples using an unbinned extended
maximum-likelihood fit to extract the \CP-violating parameters $x^{(*)}_\mp$ and $y^{(*)}_\mp$ along
with the signal and background yields.
The fit uses \mes, \DeltaE, \fis, and $m^2_\mp$ .
The likelihood for candidate $j$ is obtained by summing the product of
the event yield $N_c$, the probability density functions (PDF's) for
the kinematic and event shape variables ${\cal P}_c$, and the Dalitz
distributions ${\cal P}_c^{\rm Dalitz}$, over the signal and background
components~$c$. The likelihood function is
\bea
{\cal L} = \exp\left(-\sum_c N_{c}\right) \prod_j \sum_c N_{c}{\cal P}_{c}(\vec{\xi}_j){\cal P}^{\rm Dalitz}_{c}(\vec{\eta}_j)~,
\eea
where 
$\vec{\xi}_j = \{\mes,\DeltaE,\fis\}_j$, $\vec{\eta}_j = (m_-^2,m_+^2)_j$,
and ${\cal P}_{c}(\vec{\xi}) = {\cal P}_{1,c}(\mes) {\cal P}_{2,c}(\DeltaE) {\cal P}_{3,c}(\fis)$.
The background components in the fit are continuum, \BB and $B^-\to D^0\pi^-$ (for $B^-\to D^0 K^-$) or  
$B^-\to D^{*0}\pi^-$ (for $B^-\to D^{*0} K^-$).
For signal events, ${\cal P}^{\rm Dalitz}_{c}(\vec{\eta})$ is given by 
$\Gamma_{\mp}^{(*)}(\vec{\eta})$ 
multiplied
by the efficiency variations estimated using simulated signal events, where 
$\Gamma_{\mp}^{(*)}(\vec{\eta})$ is given by Eq.~(\ref{eq:ampgen1}).

The \mes and \DeltaE distributions for signal events are described by Gaussian
functions; the Fisher distribution is parameterized with two Gaussian functions
with different widths for the left and right parts of the curve ({\it bifurcated Gaussian}). 
Their parameters, along with most of the parameters describing the background distributions, 
are determined from a combined fit to the $\Bm \to D^{(*)0} \pim$ high-statistics control samples.

\subsection{Description of the background probability density functions}
The continuum background in the \mes distribution is described by a
threshold function~\cite{ref:argus} whose free parameter $\zeta$ is
determined from the $B^-\to D^{(*)0}\pi^-$ control samples. The continuum \DeltaE 
distribution is described by a first order polynomial whose slope is extracted 
from the control samples.
The shape of the background \mes distribution in generic \BB decays 
is taken from simulated events and uses a threshold function to describe the
combinatorial component plus a bifurcated Gaussian shape to parameterize
the peaking contribution. The fraction of the peaking contribution is extracted directly from the fit to the data.
The \DeltaE distribution for \BB background is taken from simulation and
parameterized with the sum of a second order polynomial and a Gaussian function
that takes into account the increase of combinatorial feed-down background at negative \DeltaE values.
The \mes distribution of $B^-\to D^{(*)0}\pi^-$ is the same as the signal, while the \DeltaE shape is parameterized with the 
same Gaussian function as the signal with an additional shift arising from the
wrong mass assignment to the prompt track, computed event by event as a function
of the prompt track momentum in laboratory frame and the CM boost.
The Fisher PDF for continuum background is determined from the \mes
sideband region of the control sample events and is parameterized with the sum of two Gaussian functions. The
Fisher PDF for \BB events and $B^-\to D^{(*)0}\pi^-$ background is taken to be the same as that for the signal, consistent
with the simulation.

Background events arising from continuum and \BB where the \Dz candidate is real
can mimic either the $b\to c$ or the $b\to u$ signal component, depending on 
whether the \Dz candidate is combined with a negatively or positively-charged kaon.
We take this effect into account in the likelihood function with two
parameters, the fraction $f_{D^0}$ of background events with a real
$D^0$ and the fraction $R$ of background events with a real $D^0$
associated with a negatively-charged kaon (same charge
correlation as the $b\to c$ signal component). 
These fractions have been estimated separately for continuum and \BB backgrounds from simulated events. As a check of the reliability of these estimates, the fraction $f_{D^0}$ for all background events (mixture of continuum and \BB) has been measured on data from the invariant mass distribution of $D^0$ after removing the requirement on the $D^0$ mass and using events satisfying $\mes<5.272$\gevcc. The measured value is consistent with the fraction found on simulated events. The fractions $f_{D^0}$ and $R$ for continuum and \BB background are reported in Table~\ref{tab:D0fractions}.

The shape of the Dalitz plot distribution of the continuum and
$B\bar{B}$ background is parameterized by a third-order polynomial
function in ($m_-^2$,$m_+^2$) for the combinatorial component (fake
neutral $D$ mesons), and as signal $D^0$ or $\bar{D}^0$ shapes for
real neutral $D$ mesons. 
The combinatorial distributions are taken from simulated events. The shapes for events in the $D^0$ invariant mass and \mes sidebands on data and simulated events are found to be consistent. 
The fraction of background originating from signal $B^-\to\tilde{D}^{(*)0}K^-$ 
where the $\tilde{D}^{(*)0}$ meson is combined with a combinatorial (either opposite- or same-charged) kaon
from the other $B$ meson is found to be negligible.

\begin{table}[htb]
\begin{center}
\begin{tabular}{c|c|c|c}
\hline
&&&\\[-0.4cm]
 $D^0$ fraction    &  $B^-\to \tilde{D}^{0}K^-$   &  $B^-\to \tilde{D}^{*0}(\Dztilde\pi^0)K^-$  & $B^-\to \tilde{D}^{*0}(\Dztilde\gamma)K^-$ \\  \hline
 $f_{D^0}$ (continuum)     &  $0.022\pm 0.010$  & $0.336\pm 0.038$  & $0.163\pm 0.016 $ \\  
 $R$ (continuum)           &  $0.164\pm 0.018$  & $0.170\pm 0.052$   & $0.099\pm 0.031$  \\  
 $f_{D^0}$ (\BB)    &  $0.026\pm 0.008$  & $0.130\pm 0.041$    & $0.152\pm 0.024$ \\
 $R$ (\BB)          &  $0.64\pm 0.15$  & $0.5\pm 0.5$    & $0.943\pm 0.039$ \\ \hline
\end{tabular}
\end{center}
\caption{\Dz fractions $f_{\Dz}$ and $R$, as described in the text, from simulated continuum and \BB background events.}
\label{tab:D0fractions}
\end{table}
 
\subsection{\CP\ parameters}
\label{ref:CPparams}
\begin{figure}[p]
\begin{center}
\begin{tabular}{cc}
\includegraphics[height=6.5cm]{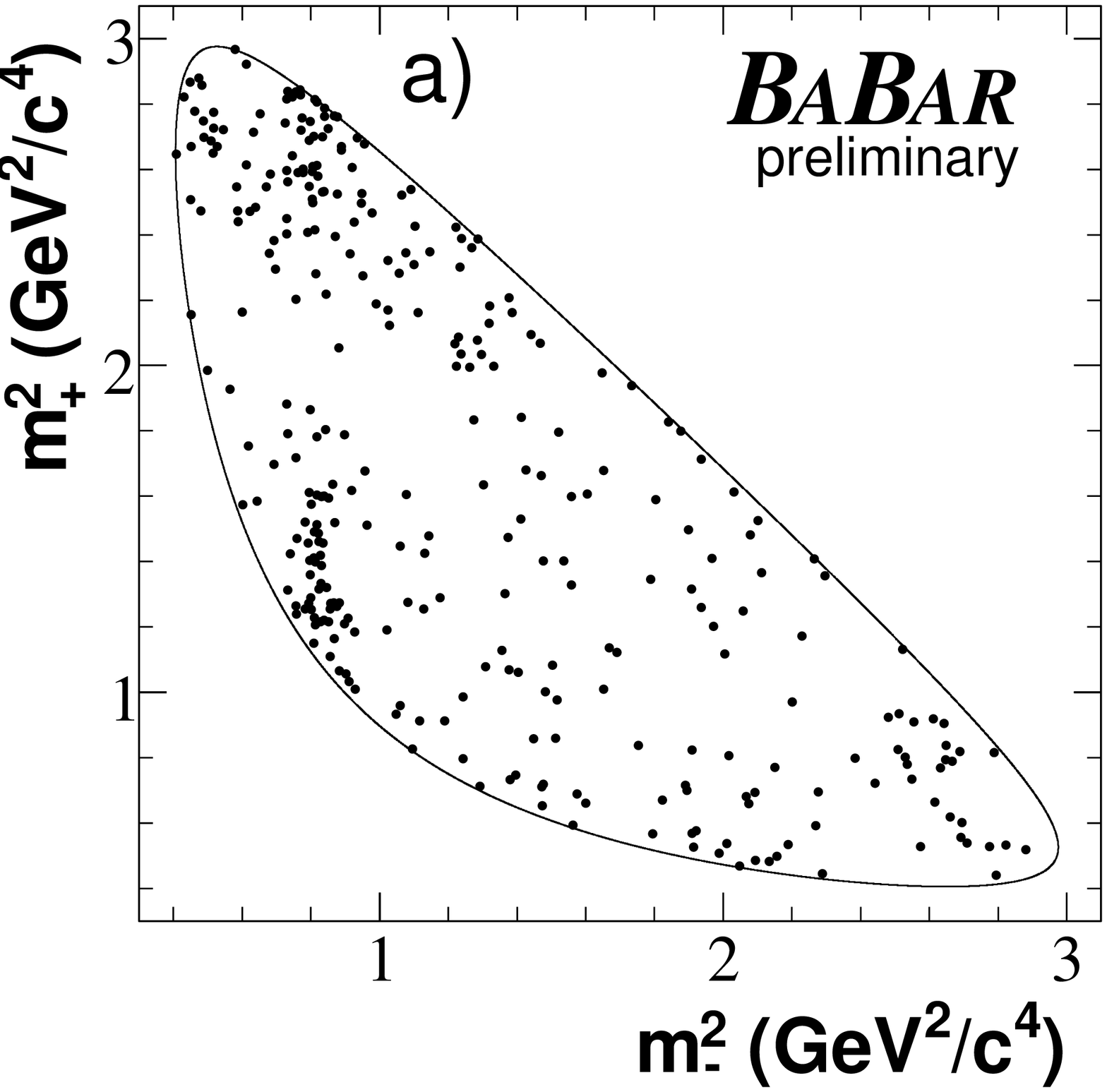}&
\includegraphics[height=6.5cm]{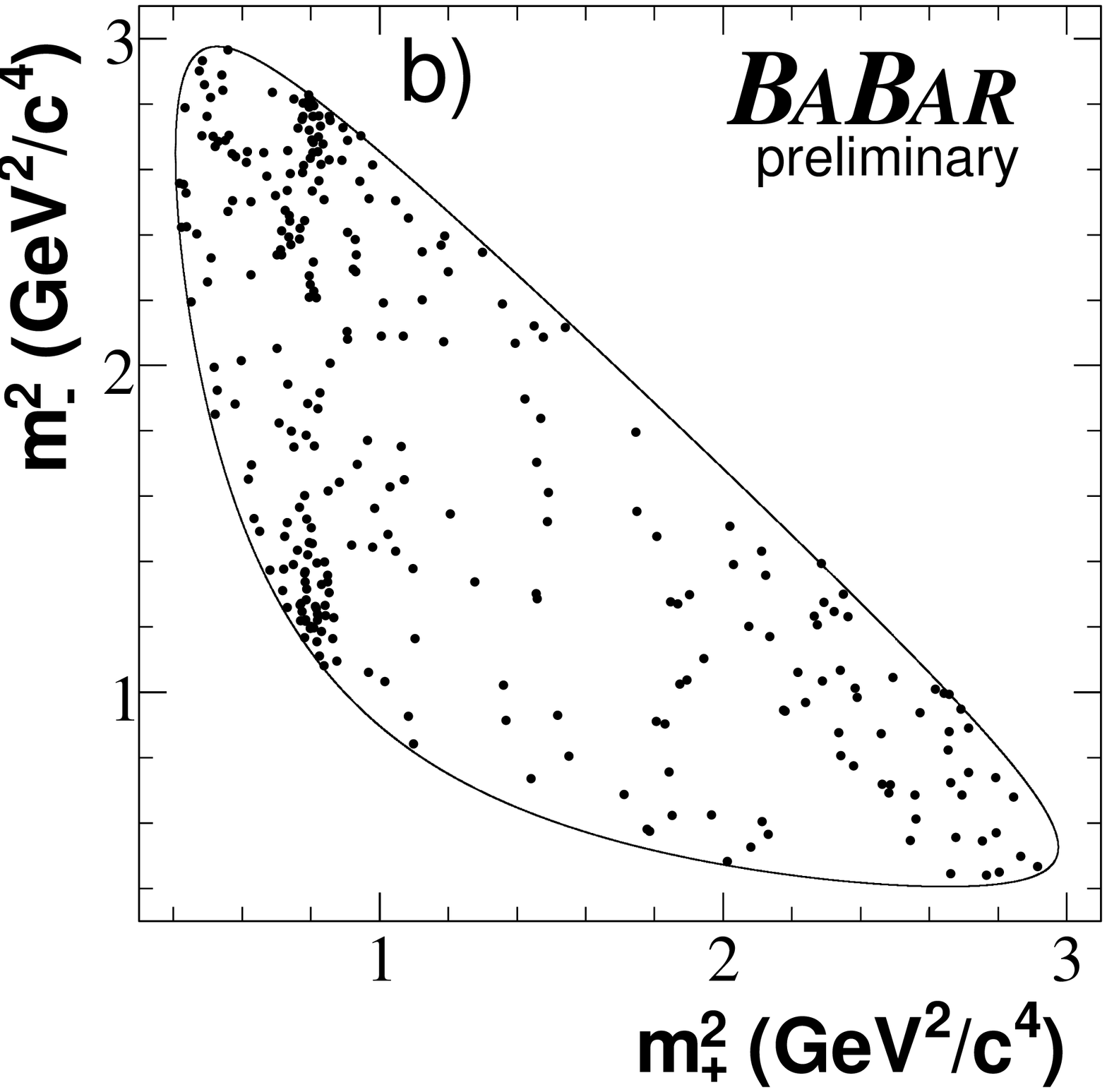}\\
\includegraphics[height=6.5cm]{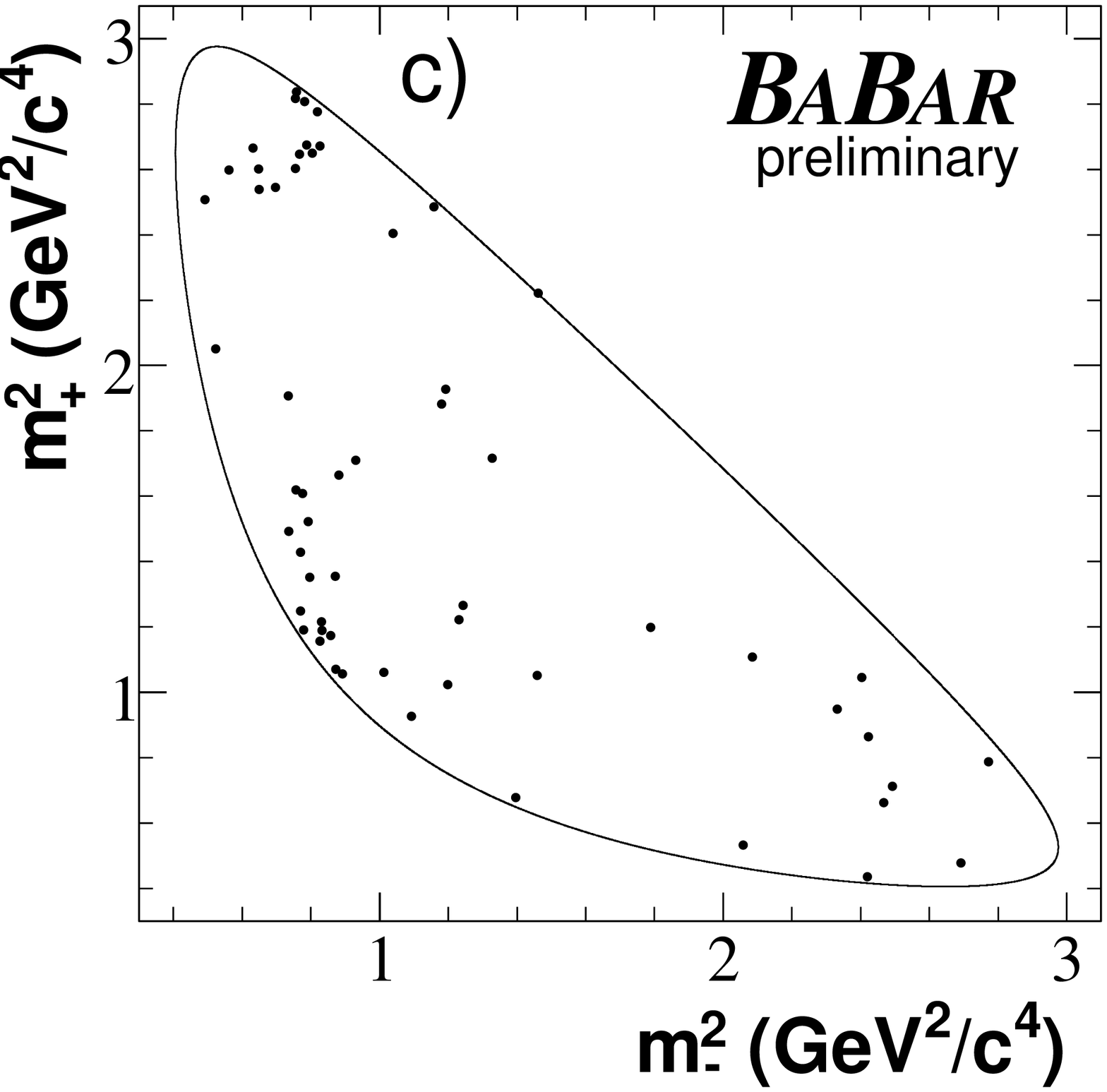}&
\includegraphics[height=6.5cm]{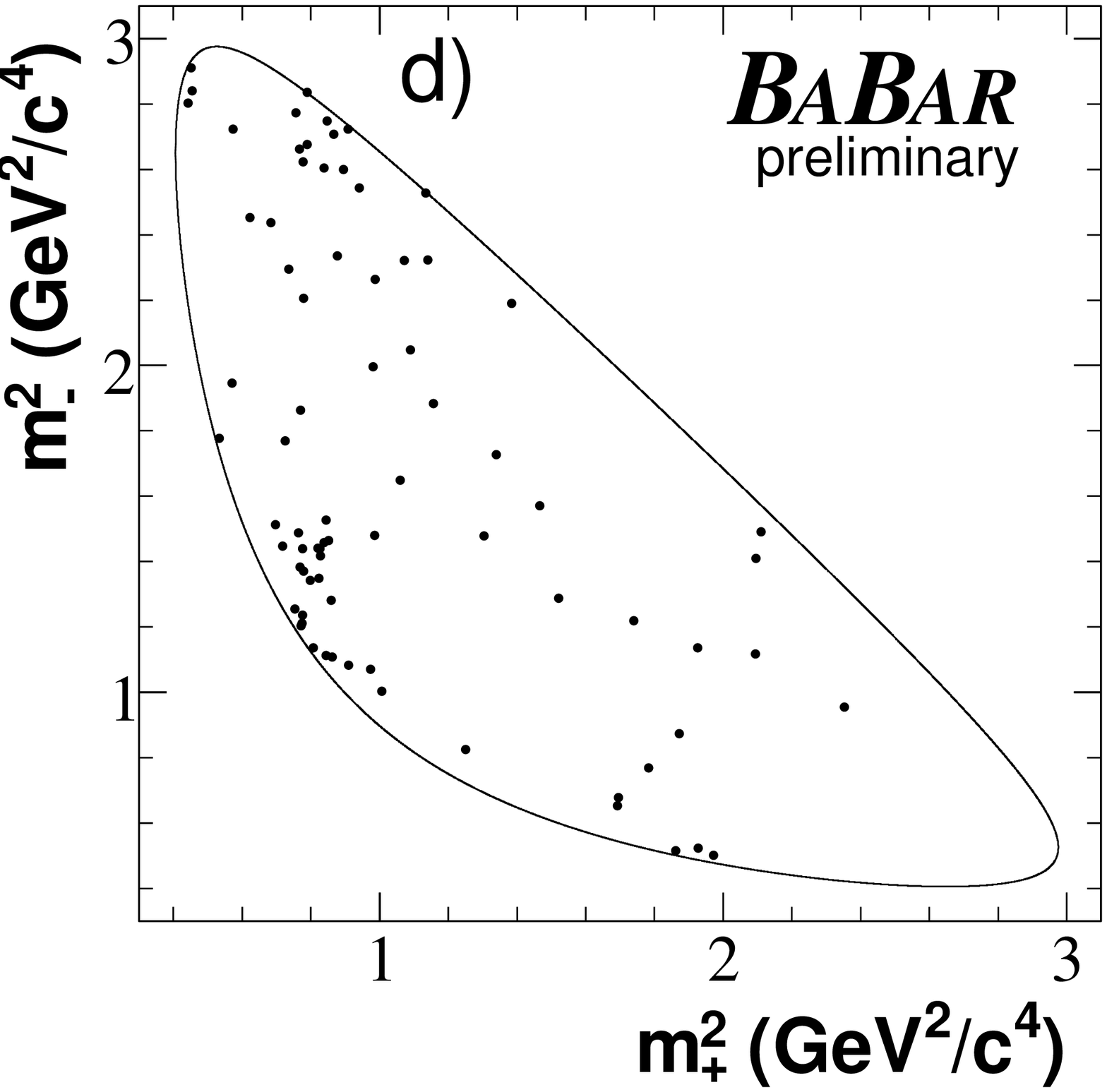}\\
\includegraphics[height=6.5cm]{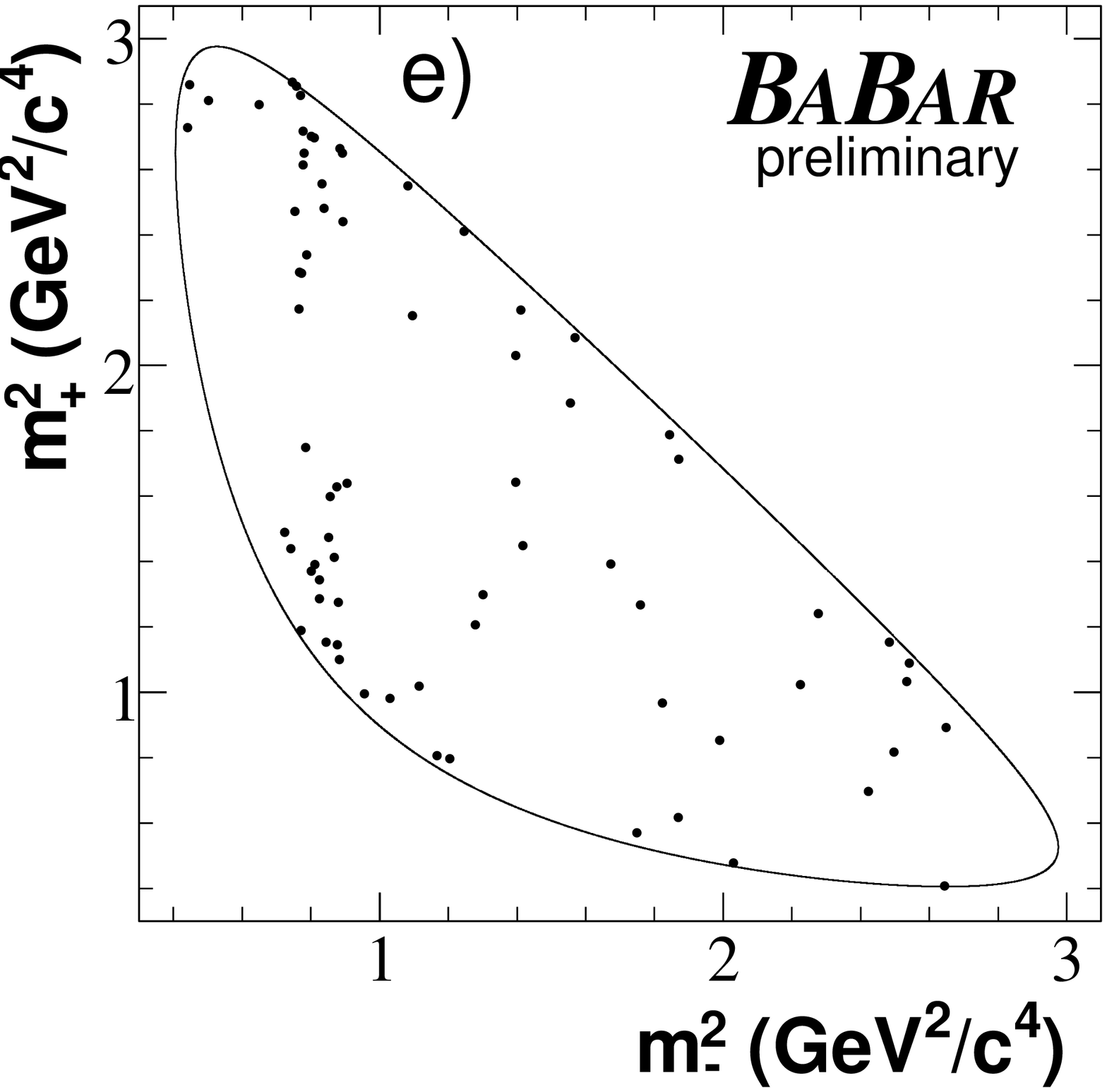}&
\includegraphics[height=6.5cm]{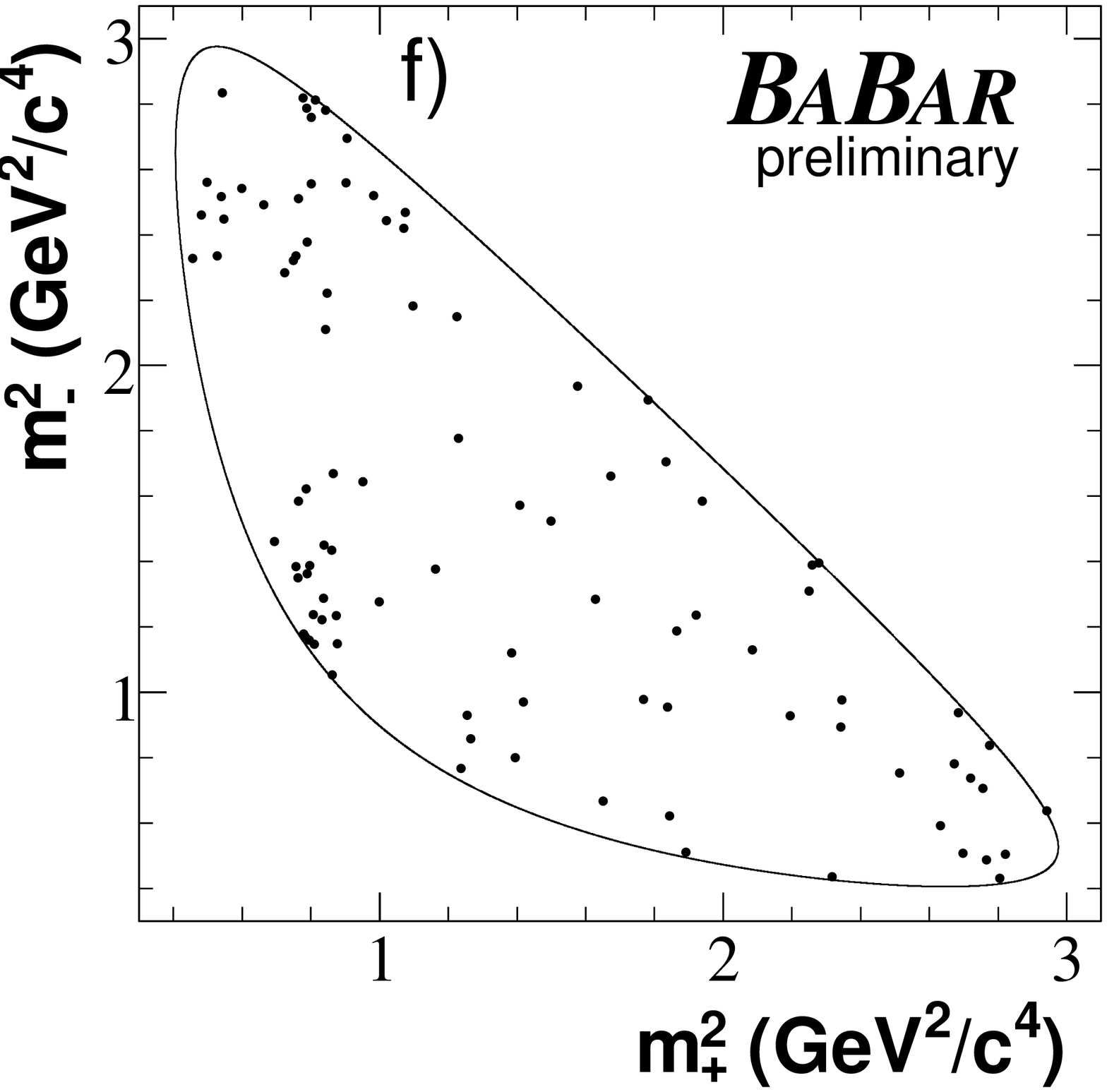}\\
\end{tabular}
\caption{The $\Dztilde\to\KS\pi^-\pi^+$ Dalitz distributions for $\Bmp \to \Dztilde \Kmp$ (a,b), 
$\Bmp \to\tilde{D}^{*0}(\Dztilde\pi^0)\Kmp$ (c,d), and
$\Bmp \to\tilde{D}^{*0}(\Dztilde\gamma)\Kmp$ (e,f), 
separately for $\Bm$ (a,c,e) and $\Bp$ (b,d,f).
The requirements $\mes>5.272$~\gevcc and $|\DeltaE|<30$~\mev have been applied to reduce the background
contamination.}
\label{fig:dalitz_dist}
\end{center}
\end{figure}

\begin{figure}[htb!]
\begin{center}
\begin{tabular}{cc}
\includegraphics[height=9cm]{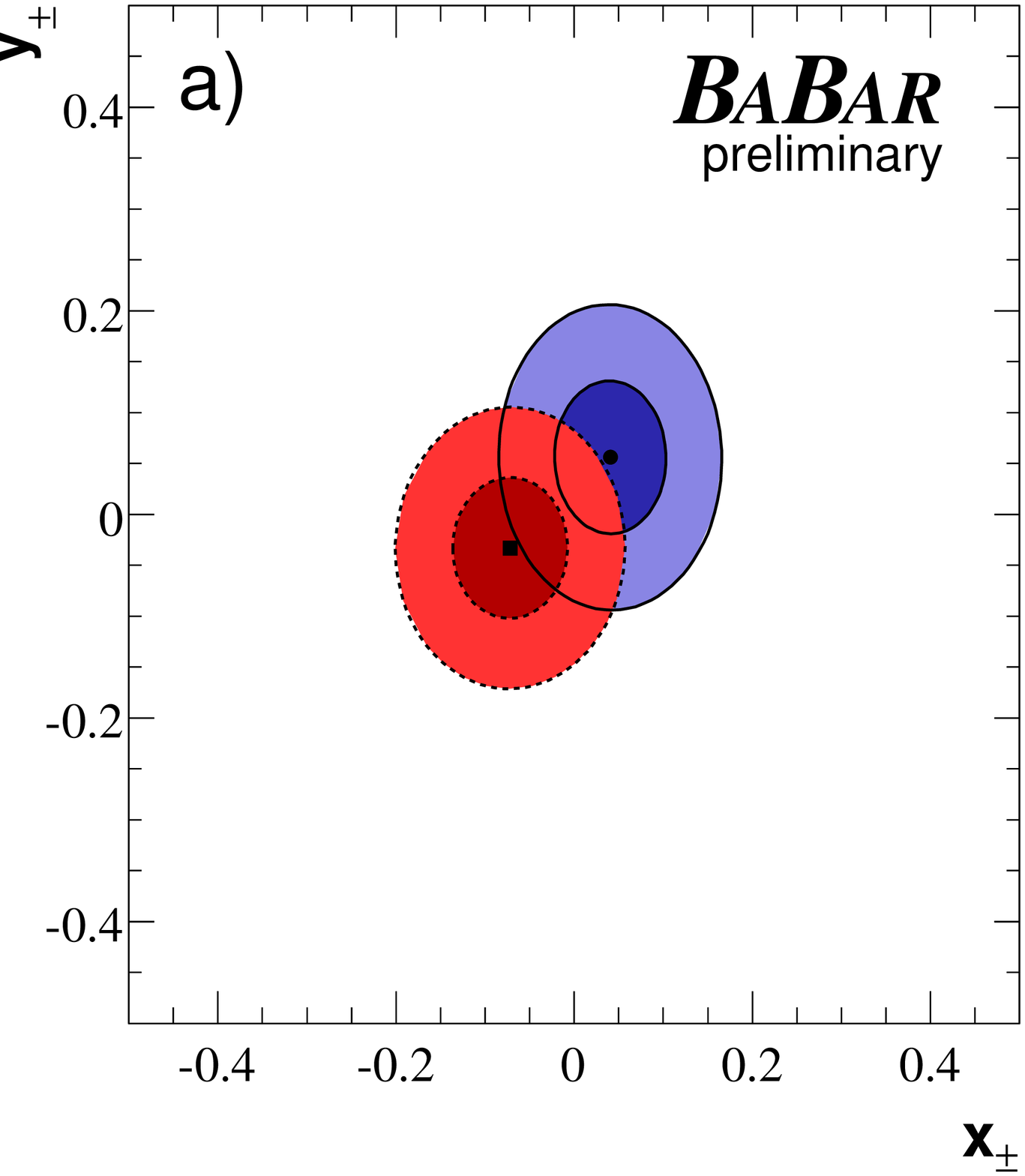}&
\includegraphics[height=9cm]{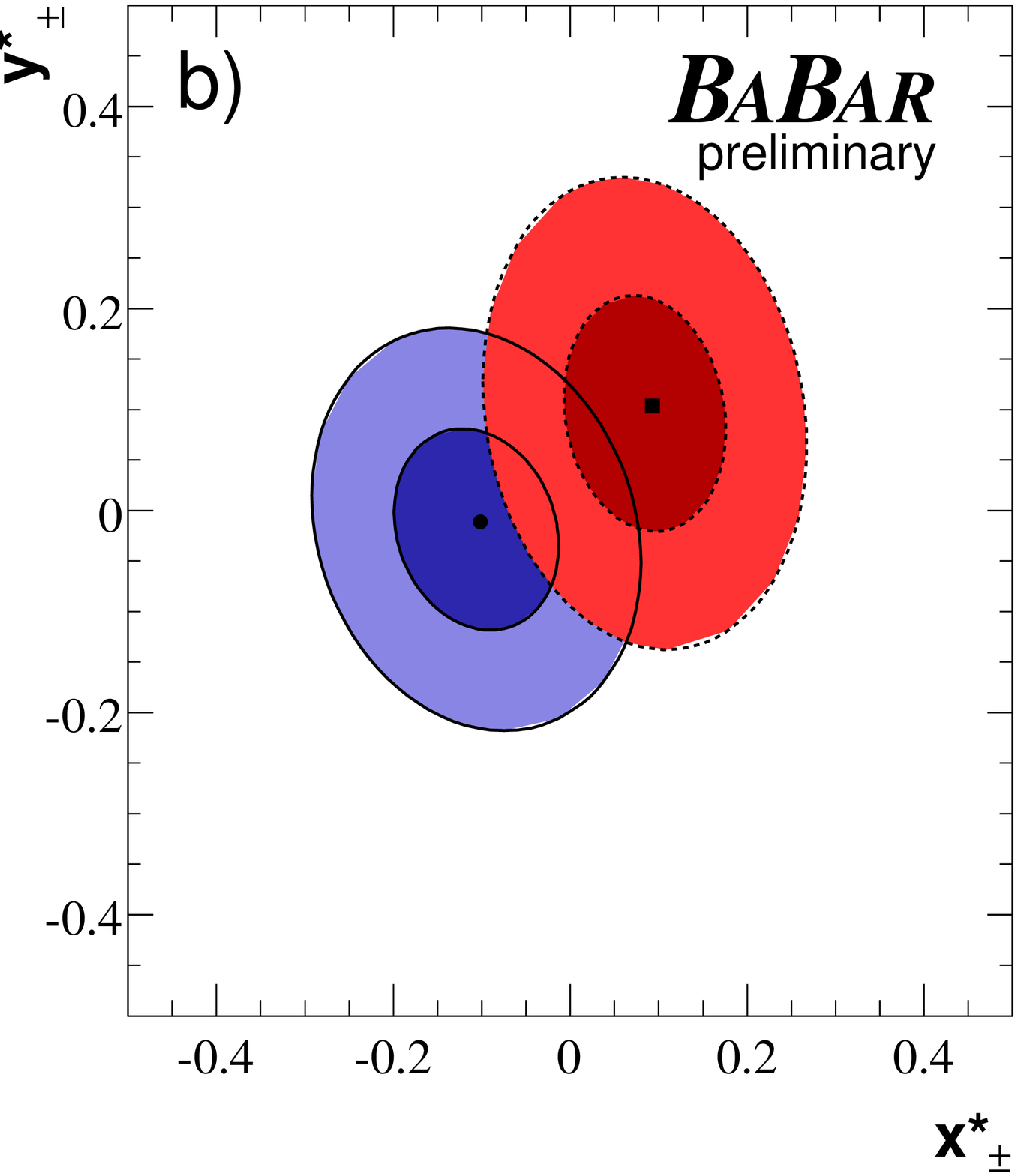}\\
\end{tabular}
\caption{Contours at $39.3\%$ (dark) and $86.5\%$ (light) confidence level (corresponding to two-dimensional one- and two-standard
deviaton regions), including statistical and systematic uncertainties, for the $(x^{(*)}_\mp,y^{(*)}_\mp)$ parameters
for \Bm (thick and solid lines) and \Bp (thin and dotted lines) decays.}
\label{fig:cart_CL}
\end{center}
\end{figure}

The signal yields measured with the \CP\ fit on the sample of $347$ million \BB events are 
$N(\Bmp \to \Dztilde \Kmp)=398\pm 23$, $N(\Bmp \to \tilde{D}^{*0}(\Dztilde\pi^0) \Kmp)=97\pm 13$, 
$N(\Bmp \to\tilde{D}^{*0}(\Dztilde\gamma) \Kmp)=93\pm 12$, 
and are consistent with expectations based on measured branching fractions and efficiencies estimated from Monte Carlo simulation. The results for the \CP-violating parameters $x^{(*)}_\mp$, $y^{(*)}_\mp$ are summarized in Table~\ref{tab:cp_coord}. 
The only non-zero statistical correlations involving the \CP parameters are 
for the pairs $(x_-,y_-)$, $(x_+,y_+)$, $(x_-^*,y_-^*)$, and $(x_+^*,y_+^*)$,
which amount to $-1\%$, $1\%$, $-17\%$, and $-14\%$, respectively.
The Dalitz plot distributions for the events selected with $\mes>5.272$\gevcc are shown in Fig.~\ref{fig:dalitz_dist} separately for $B^-$ and $B^+$ candidates. Fig.~\ref{fig:cart_CL} shows the one- and two-standard deviation confidence-level contours (including statistical and systematic uncertainties) in the $x^{(*)}-y^{(*)}$ planes for all the reconstructed modes, and separately for $B^-$ and $B^+$. The separation of the $(x^{(*)}_-,y^{(*)}_-)$ and $(x^{(*)}_+,y^{(*)}_+)$ confidence contours in these planes 
is an indication of direct \CP\ violation.

\begin{table}[htb]
\begin{center}
\begin{tabular}{c|c}
\hline 
\\[-0.15in]
\CP\ parameter    & $\Bmp \to \tilde{D}^{(*)0} \Kmp$ \\ [0.01in] \hline
$x_-$   & $\phantom{-}0.041\pm 0.059 \pm 0.018\pm 0.011$ \\ 
$y_-$   & $\phantom{-}0.056\pm 0.071\pm 0.007\pm 0.023$ \\ 
$x_+$   & $-0.072\pm 0.056 \pm 0.014\pm 0.029$ \\
$y_+$   & $-0.033\pm 0.066\pm 0.007\pm 0.018$ \\
$x_-^*$ & $-0.106\pm 0.091\pm 0.020\pm 0.009$ \\ 
$y_-^*$ & $-0.019\pm 0.096\pm 0.022\pm 0.016$ \\ 
$x_+^*$ & $\phantom{-}0.084\pm 0.088\pm 0.015\pm 0.018$ \\
$y_+^*$ & $\phantom{-}0.096\pm 0.111\pm 0.032\pm 0.017$ \\
\hline
\end{tabular}
\end{center}
\caption{\CP-violating parameters $x^{(*)}_\mp$, $y^{(*)}_\mp$ obtained from the \CP\ fit to the
 $\Bmp \to \tilde{D}^{(*)0} \Kmp$ samples. The first error is statistical,
 the second is experimental systematic uncertainty and the third is
 the systematic uncertainty associated with the Dalitz model.}
\label{tab:cp_coord}
\end{table}

\subsection{Systematic error associated with the $D^0$ Dalitz model}\label{sec:syst_model}
The largest single contribution to the systematic uncertainties in the \CP\ parameters comes from the choice of the Dalitz model used to describe the $D^0\to\KS\pi^-\pi^+$ decay amplitude. The $D^0$ sample used to determine the reference model introduced in Sec.~\ref{sec:dalitzmodel} is fitted with a set of alternative models where the resonances are described with different parameterizations or removed:
\begin{itemize}
\item[1)] $\pi\pi$ S-wave: the reference model uses two wide BW scalar amplitudes ($\sigma$ and $\sigma'$). Alternatively we use a 
K-matrix model~\cite{ref:babar_dalitzeps05} with pole masses and coupling constants fixed by fits to scattering data~\cite{ref:AS}. 
See also Sec.~\ref{sec:dalitzmodel}. 
\item[2)] $\pi\pi$ P-wave: the mass and the width of the Gounaris-Sakurai BW describing the $\rho(770)$ are changed within their quoted uncertainty~\cite{ref:pdg2004}.
\item[3)] $\pi\pi$ and $K\pi$ D-waves: alternative to the helicity formalism used in the reference model, for $f_2(1270)$ and $K^*_2(1430)$ we use the formalism derived from Zemach tensors~\cite{ref:zemach}. The difference is very small for P-waves but is larger for D-waves.  
\item[4)] $K\pi$ S-wave: the mass and width of the BW
  describing $K^*(1430)$ are taken from
  E791~\cite{ref:e791K*}. Alternatively, we have floated them in our
  flavor tagged $D^0$ sample obtaining consistent values. As an
  additional model we use an adaptation of the LASS 
parameterization~\cite{ref:lassparam1} with parameters taken from the fit to our $\Dstarp\to\Dz\pip$ data sample. 
\item[5)] $K\pi$ P-wave: it is dominated by the $K^*(892)$ in both
  Cabibbo allowed and doubly Cabibbo suppressed amplitude. The mass
  and the width of this resonance, taken from PDG~\cite{ref:pdg2004}
  in the reference model, are changed to the values found by keeping
  them floating in the fit to the flavour-tagged $D^0$ sample. The
  resulting values are consistent with what is found in $B\to J/\Psi
  K\pi$ decays selected in \babar\ data.
\item[6)] Blatt-Weisskopf penetration factors: the effect from the Blatt-Weisskopf penetration factors has been evaluated using an alternative model that doesn't include 
them~\cite{ref:blatt}.
\item[7)] Running width of BW: a model with BW's of fixed width is used.
\item[8)] $K^*_2(1430)$, $K^*(1680)$, $K^*(1410)$ and $\rho(1450)$: these resonances are removed from the reference model.
\end{itemize}
We have generated a sample of $B^\mp\to \tilde{D}^0 K^\mp$ and $B^\mp\to \tilde{D}^{*0}K^\mp$ signal
events that is one hundred times larger than the measured signal yields in data. 
The Dalitz plot distribution of $D^0$ is generated according to the reference
model and to \CP\ parameters consistent with the values found in
data. The \CP\ parameters are extracted by fitting the generated
Dalitz plot distributions using a PDF equal to the reference model
({\em model 0}) or to one of the eight alternative models ({\it model 1, 2,...,8}). We take
as the systematic uncertainty of $(x_\mp,y_\mp)$ --- similarly for
$(x_\mp^*,y_\mp^*)$ --- associated with the
$i^{th}$ alternative model the difference between the \CP\ parameters
fitted using the alternative model ($x^i_\mp,y^i_\mp$) and the reference model ($x^0_\mp,y^0_\mp$):
$\Delta x_\mp^i= x_\mp^i-x_\mp^0$, $\Delta y_\mp^i=
y_\mp^i-y_\mp^0$. As total systematic uncertainty associated with the
Dalitz model we consider the sum square of contributions from the
alternative models: $\Delta x_\mp=\sqrt{\sum_{i=1}^8 {\Delta
    x_\mp^i}^2}$,  $\Delta y_\mp=\sqrt{\sum_{i=1}^8 {\Delta
    y_\mp^i}^2}$. The dominant contributions to the overall
Dalitz model uncertainty arise from the models 1), 4), and 7).
The systematic uncertainties associated with the Dalitz model are summarized in Table~\ref{tab:cartesian-syst}.

\begin{table}[htpb]
\begin{center}
\begin{tabular}{l||c|c|c|c||c|c|c|c}
\hline
\\[-0.15in]
 Source                               &$x_-$  & $y_-$ & $x_+$ & $y_+$  & $x^*_-$ & $y^*_-$ & $x^*_+$ & $y^*_+$  \\   [0.01in] \hline
 \mes, \DeltaE, \fis shapes           & 0.002 & 0.004 & 0.003 & 0.004  & 0.011 & 0.012 & 0.008 & 0.008   \\
 Real \Dz\ fractions                  & 0.002 & 0.000 & 0.000 & 0.000  & 0.002 & 0.003 & 0.002 & 0.016   \\
 Fraction of right sign \Dz's         & 0.008 & 0.002 & 0.002 & 0.002  & 0.005 & 0.005 & 0.001 & 0.022  \\
 Efficiency in the Dalitz plot        & 0.014 & 0.000 & 0.013 & 0.001  & 0.001 & 0.002 & 0.000 & 0.001   \\
 Background Dalitz shape              & 0.006 & 0.003 & 0.001 & 0.004  & 0.012 & 0.015 & 0.009 & 0.009  \\
 Dalitz amplitudes and phases         & 0.004 & 0.004 & 0.004 & 0.004  & 0.008 & 0.008 & 0.008 & 0.008  \\
 $B^-\to D^{*0}K^-$ cross-feed        & 0.000 & 0.000 & 0.000 & 0.000  & 0.004 & 0.001 & 0.004 & 0.004        \\
 \CP violation in $D\pi$ and \BB bkg  & 0.000 & 0.000 & 0.000 & 0.000  & 0.005 & 0.002 & 0.002 & 0.005   \\ \hline
 Total experimental                   & 0.018 & 0.007 & 0.014 & 0.007  & 0.020 & 0.022 & 0.015 & 0.032   \\ 
 $\Dz$ Dalitz model                   & 0.011 & 0.023 & 0.029 & 0.018  & 0.009 & 0.016 & 0.018 & 0.017 \\ \hline 
 Total                                & 0.021 & 0.024 & 0.032 & 0.019  & 0.021 & 0.027 & 0.023 & 0.036 \\ \hline 
\end{tabular}
\end{center}
\caption{Summary of the main contributions to the systematic error on the \CP parameters $x_\mp$, $y_\mp$, $x^*_\mp$, and $y^*_\mp$.
\label{tab:cartesian-syst}}
\end{table}

\subsection{Experimental systematic errors}

The main experimental systematic errors are listed in Table~\ref{tab:cartesian-syst}.
Uncertainties due to the \mes, \DeltaE, and \fis\ PDF parameters for signal and background
extracted from the combined fit to the $\Bm \to D^{(*)0} \pim$ control samples (fixed in the 
reference \CP fit) are estimated from the statistical differences on $x^{(*)}_\mp$ and $y^{(*)}_\mp$ when 
the former set of parameters is also floated in the \CP fit. 
Other \mes, \DeltaE, and \fis\ parameters fixed in the \CP fit are changed by one standard deviation. The uncertainties 
associated to the knowledge of the fraction of background events with a real $D^0$ and the Dalitz
distribution of background events are evaluated from the differences on the \CP parameters 
when the estimates obtained from simulated events are replaced by the estimates using sideband data. 
The systematic uncertainty on the fraction of events where a true \Dz is associated with a
negatively-charged kaon is obtained from the variation of the \CP parameters when
the \Dz is randomly associated either to a negatively- or positively-charged kaon (absence of charge correlation).
The effect due to reconstruction efficiency variations of the signal across the Dalitz plane has been estimated
assuming a perfectly uniform efficiency. The statistical errors in the Dalitz amplitudes and phases from the fit to the 
tagged $D^0$ sample have been propagated to the $x^{(*)}_\mp$ and $y^{(*)}_\mp$ parameters performing
a simultaneous \CP and Dalitz fit to the $\Bm \to D^{(*)0} \Km$ and $\Dstarp\to\Dz\pip$ data.
The effect of the remaining cross-feed of $B^-\to\tilde{D}^{*0}(D^0\pi^0)K^-$ events into the 
$B^-\to\tilde{D}^{*0}(D^0\gamma)K^-$ sample (5\% of the signal yield) has been evaluated by including an additional background 
component with ${\cal P}^{\rm Dalitz}_{c}(\vec{\eta})$ identical to that of $B^-\to\tilde{D}^{*0}(D^0\piz)K^-$ signal events.
Finally, possible \CP-violating effects in the background
have been evaluated by setting the \CP parameters of the $B^-\to D^{(*)0}\pi^-$
 background component to the values obtained 
from a \CP fit to the $\Bm \to D^{(*)0} \pim$ control samples, and by floating an independent set of \CP parameters
for the other \BB background.

The following sources of uncertainty are found to be negligible: 
the assumption of perfect mass resolution for
the Dalitz plot variables $(m^2_-,m^2_+)$, the presence of combinatorial
background from signal events where the prompt kaon is replaced by a combinatorial track,
and the assumption that the shape of the
continuum or \BB background does not change when the $D^0$ is fake or real.

\section{INTERPRETATION}\label{sec:results}
\begin{figure}[htb!]
\begin{center}
\begin{tabular}{cc}
\includegraphics[height=8cm]{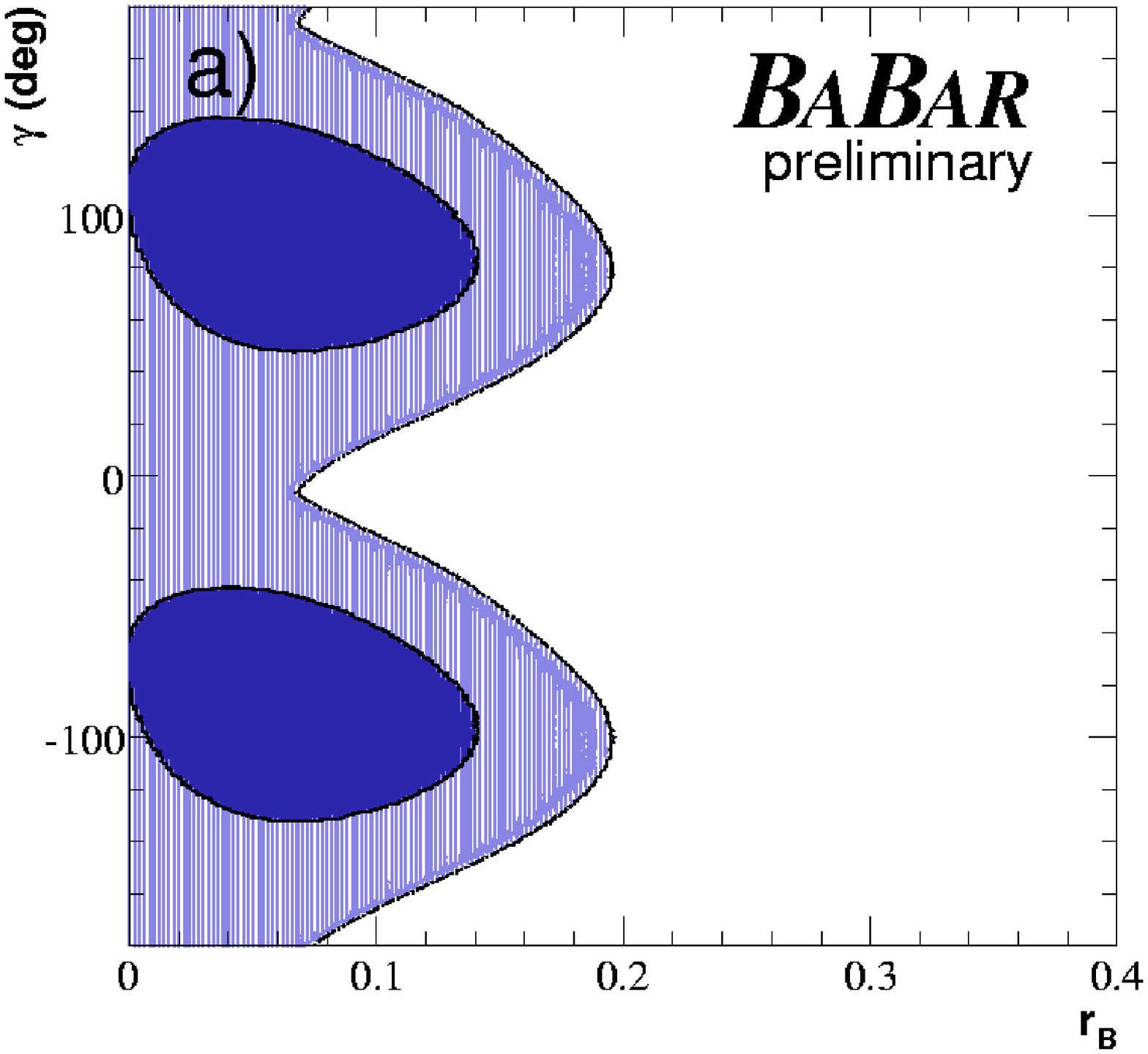}&
\includegraphics[height=8cm]{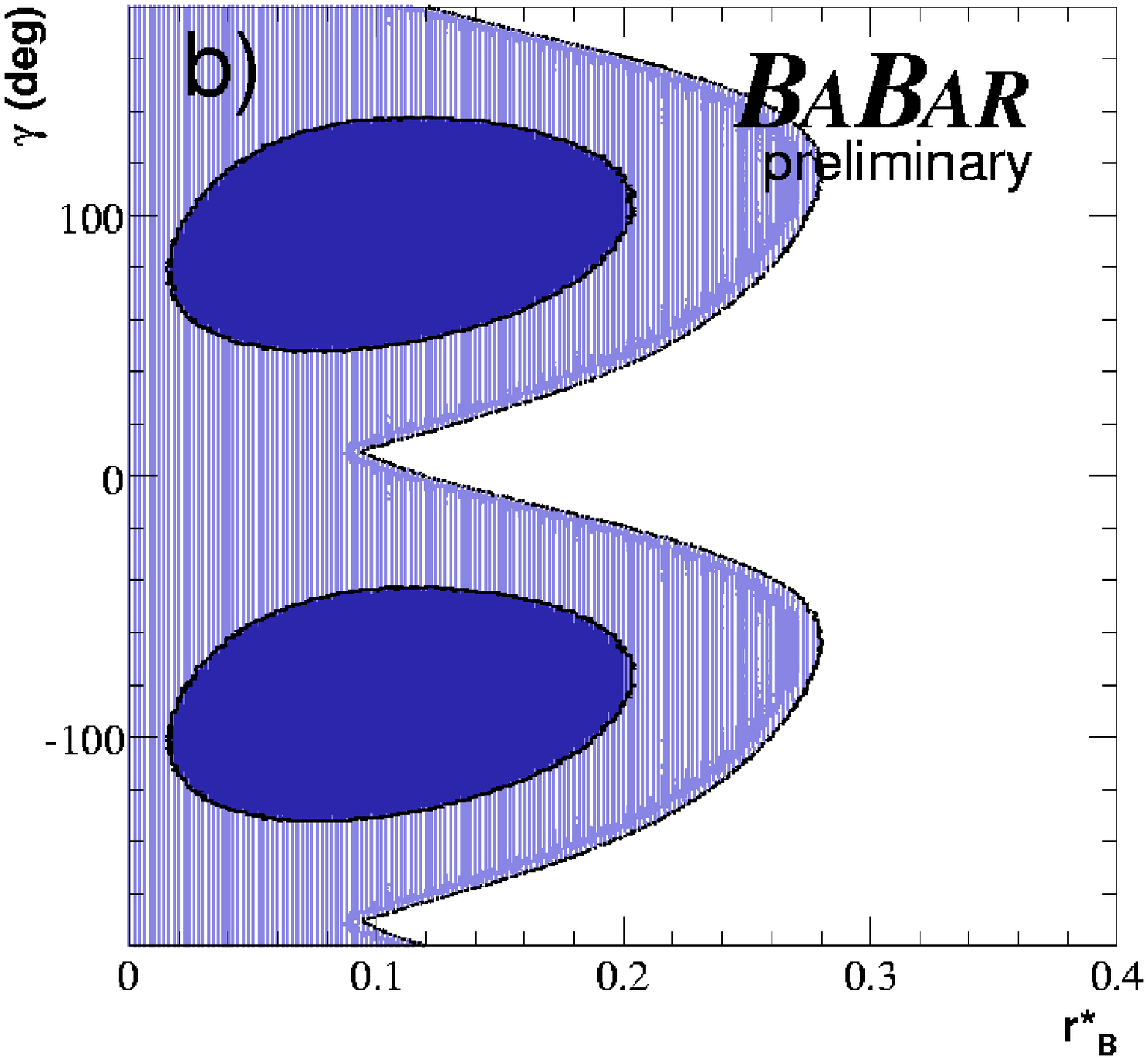}\\
\end{tabular}
\caption{Projections in the (a) $r_B-\gamma$ and (b) $r^*_B-\gamma$
  planes of the five-dimensional one- (dark) and two- (light) standard deviation regions.}
\label{fig:gamma_rb_proj}
\end{center}
\end{figure}

A frequentist (Neyman) procedure~\cite{ref:pdg2004,ref:neyman} identical to that used in our previous
measurements~\cite{ref:babar_dalitzpub,ref:babar_dalitzeps05} has been adopted to interpret
the measurement of the \CP\ parameters $(x^{(*)}_\mp,y^{(*)}_\mp)$ 
reported in table~\ref{tab:cp_coord} 
in terms of confidence regions on ${\bf p}=(\gamma,r_B,\delta_B,r^*_B,\delta^*_B)$.
Using a large number of pseudo-experiments with probability density functions
and parameters as obtained from the fit to the data but with many different values of the 
\CP\ parameters, we construct 
a multivariate Gaussian 
parameterization of
the PDF of $(x^{(*)}_\mp,y^{(*)}_\mp)$ as a function of ${\bf p}$
which takes into account the statistical and systematic correlations.
For a given ${\bf p}$, the five-dimensional confidence level ${\cal C}=1-\alpha$
is calculated by integrating over all points in the fit parameter
space closer (larger PDF) to ${\bf p}$ than the fitted data values. 
The one- (two-) standard deviation region of the \CP
parameters is defined as the set of ${\bf p}$ values for which
$\alpha$ is smaller than 3.7\% (45.1\%). 
Figure~\ref{fig:gamma_rb_proj} shows the two-dimensional
projections onto the $r_B-\gamma$ and $r^*_B-\gamma$ planes, including
statistical and systematic uncertainties. 
The figure shows that this Dalitz analysis has a two-fold ambiguity,
$(\gamma,\delta^{(*)}_B) \to (\gamma + 180^\circ, \delta^{(*)}_B +
180^\circ)$, as expected from Eq.~(\ref{eq:ampgen1}).
From the one-dimensional projections we obtain for the weak phase
$\gamma = (92 \pm 41 \pm 11 \pm 12)^\circ$,
and for the strong phase differences  
$\delta_{\B} = (118 \pm 63 \pm 19 \pm 36)^\circ$ and 
$\delta^*_{\B} = (-62 \pm 59 \pm 18 \pm 10)^\circ$. No constraints on the phases are achieved at two
standard deviation level and beyond. 
Similarly, for the
magnitude of the ratio of decay amplitudes $r_\B$ and $r_\B^*$ we obtain
the one (two) standard deviations 
constraints $r_\B<0.140~(r_\B<0.195)$ and $0.017<r_\B^*<0.203~(r_\B^*<0.279)$.
All these results are obtained considering the statistical correlations mentioned in Sec.~\ref{ref:CPparams},
while the experimental and Dalitz model systematic uncertainties are taken uncorrelated. 
We have verified that accounting for experimental systematic correlations within
a given measurement $(x_\mp,y_\mp)$ or $(x^*_\mp,y^*_\mp)$, or assuming 
the experimental and Dalitz model systematic uncertainties between
$(x_\mp,y_\mp)$ and $(x^*_\mp,y^*_\mp)$ fully correlated, has a negligible effect on the results.

\section{CONCLUSIONS}
We have presented a preliminary updated measurement of the \CP\ parameters
$(x_\mp,y_\mp)$ and $(x^*_\mp,y^*_\mp)$ with 
$B^\mp\to \tilde{D}^{(*)0}K^\mp$, 
$\tilde{D}^{*0} \rightarrow \Dztilde\pi^0,\Dztilde\gamma$,
$\tilde{D}^0\to \KS\pi^-\pi^+$ decays based on a data sample of 347 million
\BB pairs, that supersedes the previous one based on
about 227 million \BB 
pairs~\cite{ref:babar_dalitzpub}. 
The current analysis reduces the experimental systematic uncertainty and improves the procedure to estimate the error associated 
with the Dalitz model of the $D^0$ decay.

Despite the improved measurement of $(x^{(*)}_\mp,y^{(*)}_\mp)$, the uncertainty on $\gamma$ has 
increased with respect to our previous measurement~\cite{ref:babar_dalitzpub}, moving
from $\gamma = (70\pm31^{+12+14}_{-10-11})^\circ$
to $\gamma = (92 \pm 41 \pm 11 \pm 12)^\circ$.
Since the uncertainty on $\gamma$ scales roughly as $1/r^{(*)}_B$,
this change is explained by noticing that the new $(x^{(*)}_\mp,y^{(*)}_\mp)$ measurement 
favors values of $r^{(*)}_B$ smaller than our previous analysis and
significantly smaller than the latest Belle results~\cite{ref:belle_dalitz}.

\section{ACKNOWLEDGMENTS}
\label{sec:Acknowledgments}

We are grateful for the 
extraordinary contributions of our \pep2\ colleagues in
achieving the excellent luminosity and machine conditions
that have made this work possible.
The success of this project also relies critically on the 
expertise and dedication of the computing organizations that 
support \babar.
The collaborating institutions wish to thank 
SLAC for its support and the kind hospitality extended to them. 
This work is supported by the
US Department of Energy
and National Science Foundation, the
Natural Sciences and Engineering Research Council (Canada),
Institute of High Energy Physics (China), the
Commissariat \`a l'Energie Atomique and
Institut National de Physique Nucl\'eaire et de Physique des Particules
(France), the
Bundesministerium f\"ur Bildung und Forschung and
Deutsche Forschungsgemeinschaft
(Germany), the
Istituto Nazionale di Fisica Nucleare (Italy),
the Foundation for Fundamental Research on Matter (The Netherlands),
the Research Council of Norway, the
Ministry of Science and Technology of the Russian Federation, 
Ministerio de Educaci\'on y Ciencia (Spain), and the
Particle Physics and Astronomy Research Council (United Kingdom). 
Individuals have received support from 
the Marie-Curie IEF program (European Union) and
the A. P. Sloan Foundation.


\begin{thebibliography}{99}
\bibitem{ref:chargeconj} Reference to the charge-conjugate state is implied here and throughout the text unless otherwise specified.
\bibitem{ref:comment_sigma} The $\sigma$ and $\sigma'$ masses and widths are determined from the data. We find (in~\mevcc) $M_{\sigma}=490\pm6$, $\Gamma_{\sigma}=406\pm11$, 
$M_{\sigma'}=1024\pm4$, and $\Gamma_{\sigma'}=89\pm7$. 
Errors are statistical.
\end{thebibliography}
\end{document}